
\def\unlock{\catcode`@=11} 
\def\lock{\catcode`@=12} 
\unlock
%
%
%
%
%

\font\fourteenrm=cmr10 scaled\magstep2
\font\twelverm=cmr10 scaled\magstep1
\font\ninerm=cmr9          \font\sixrm=cmr6

\font\fourteenbf=cmbx10 scaled\magstep2
\font\twelvebf=cmbx10 scaled\magstep1

\font\seventeeni=cmmi10 scaled\magstep3     \skewchar\seventeeni='177
\font\fourteeni=cmmi10 scaled\magstep2      \skewchar\fourteeni='177
\font\twelvei=cmmi10 scaled\magstep1        \skewchar\twelvei='177
\font\ninei=cmmi9                           \skewchar\ninei='177
\font\sixi=cmmi6                            \skewchar\sixi='177
\font\seventeensy=cmsy10 scaled\magstep3    \skewchar\seventeensy='60
\font\fourteensy=cmsy10 scaled\magstep2     \skewchar\fourteensy='60
\font\twelvesy=cmsy10 scaled\magstep1       \skewchar\twelvesy='60
\font\ninesy=cmsy9                          \skewchar\ninesy='60
\font\sixsy=cmsy6                           \skewchar\sixsy='60

\font\fourteenex=cmex10 scaled\magstep2
\font\twelveex=cmex10 scaled\magstep1

\font\fourteensl=cmsl10 scaled\magstep2
\font\twelvesl=cmsl10 scaled\magstep1

\font\fourteenit=cmti10 scaled\magstep2
\font\twelveit=cmti10 scaled\magstep1
\font\twelvett=cmtt10 scaled\magstep1
\font\twelvecp=cmcsc10 scaled\magstep1
\font\tencp=cmcsc10
\newfam\cpfam
%
%
\newcount\f@ntkey            \f@ntkey=0
\def\samef@nt{\relax \ifcase\f@ntkey \rm \or\oldstyle \or\or
         \or\it \or\sl \or\bf \or\tt \or\caps \fi }
\def\fourteenpoint{\relax
    \textfont0=\fourteenrm          \scriptfont0=\tenrm
    \scriptscriptfont0=\sevenrm
     \def\rm{\fam0 \fourteenrm \f@ntkey=0 }\relax
    \textfont1=\fourteeni           \scriptfont1=\teni
    \scriptscriptfont1=\seveni
     \def\oldstyle{\fam1 \fourteeni\f@ntkey=1 }\relax
    \textfont2=\fourteensy          \scriptfont2=\tensy
    \scriptscriptfont2=\sevensy
    \textfont3=\fourteenex     \scriptfont3=\fourteenex
    \scriptscriptfont3=\fourteenex
    \def\it{\fam\itfam \fourteenit\f@ntkey=4 }\textfont\itfam=\fourteenit
    \def\sl{\fam\slfam \fourteensl\f@ntkey=5 }\textfont\slfam=\fourteensl
    \scriptfont\slfam=\tensl
    \def\bf{\fam\bffam \fourteenbf\f@ntkey=6 }\textfont\bffam=\fourteenbf
    \scriptfont\bffam=\tenbf     \scriptscriptfont\bffam=\sevenbf
    \def\tt{\fam\ttfam \twelvett \f@ntkey=7 }\textfont\ttfam=\twelvett
    \h@big=11.9\p@{} \h@Big=16.1\p@{} \h@bigg=20.3\p@{} \h@Bigg=24.5\p@{}
    \def\caps{\fam\cpfam \twelvecp \f@ntkey=8 }\textfont\cpfam=\twelvecp
    \setbox\strutbox=\hbox{\vrule height 12pt depth 5pt width\z@}
    \samef@nt}
\def\twelvepoint{\relax
    \textfont0=\twelverm          \scriptfont0=\ninerm
    \scriptscriptfont0=\sixrm
     \def\rm{\fam0 \twelverm \f@ntkey=0 }\relax
    \textfont1=\twelvei           \scriptfont1=\ninei
    \scriptscriptfont1=\sixi
     \def\oldstyle{\fam1 \twelvei\f@ntkey=1 }\relax
    \textfont2=\twelvesy          \scriptfont2=\ninesy
    \scriptscriptfont2=\sixsy
    \textfont3=\twelveex          \scriptfont3=\twelveex
    \scriptscriptfont3=\twelveex
    \def\it{\fam\itfam \twelveit \f@ntkey=4 }\textfont\itfam=\twelveit
    \def\sl{\fam\slfam \twelvesl \f@ntkey=5 }\textfont\slfam=\twelvesl
    \scriptfont\slfam=\ninerm
    \def\bf{\fam\bffam \twelvebf \f@ntkey=6 }\textfont\bffam=\twelvebf
    \scriptfont\bffam=\ninerm     \scriptscriptfont\bffam=\sixrm
    \def\tt{\fam\ttfam \twelvett \f@ntkey=7 }\textfont\ttfam=\twelvett
    \h@big=10.2\p@{}
    \h@Big=13.8\p@{}
    \h@bigg=17.4\p@{}
    \h@Bigg=21.0\p@{}
    \def\caps{\fam\cpfam \twelvecp \f@ntkey=8 }\textfont\cpfam=\twelvecp
    \setbox\strutbox=\hbox{\vrule height 10pt depth 4pt width\z@}
    \samef@nt}
\def\tenpoint{\relax
    \textfont0=\tenrm          \scriptfont0=\sevenrm
    \scriptscriptfont0=\fiverm
    \def\rm{\fam0 \tenrm \f@ntkey=0 }\relax
    \textfont1=\teni           \scriptfont1=\seveni
    \scriptscriptfont1=\fivei
    \def\oldstyle{\fam1 \teni \f@ntkey=1 }\relax
    \textfont2=\tensy          \scriptfont2=\sevensy
    \scriptscriptfont2=\fivesy
    \textfont3=\tenex          \scriptfont3=\tenex
    \scriptscriptfont3=\tenex
    \def\it{\fam\itfam \tenit \f@ntkey=4 }\textfont\itfam=\tenit
    \def\sl{\fam\slfam \tensl \f@ntkey=5 }\textfont\slfam=\tensl
    \def\bf{\fam\bffam \tenbf \f@ntkey=6 }\textfont\bffam=\tenbf
    \scriptfont\bffam=\sevenbf     \scriptscriptfont\bffam=\fivebf
    \def\tt{\fam\ttfam \tentt \f@ntkey=7 }\textfont\ttfam=\tentt
    \def\caps{\fam\cpfam \tencp \f@ntkey=8 }\textfont\cpfam=\tencp
    \setbox\strutbox=\hbox{\vrule height 8.5pt depth 3.5pt width\z@}
    \samef@nt}
%
%
%
%
\newdimen\h@big  \h@big=8.5\p@
\newdimen\h@Big  \h@Big=11.5\p@
\newdimen\h@bigg  \h@bigg=14.5\p@
\newdimen\h@Bigg  \h@Bigg=17.5\p@
\def\big#1{{\hbox{$\left#1\vbox to\h@big{}\right.\n@space$}}}
\def\Big#1{{\hbox{$\left#1\vbox to\h@Big{}\right.\n@space$}}}
\def\bigg#1{{\hbox{$\left#1\vbox to\h@bigg{}\right.\n@space$}}}
\def\Bigg#1{{\hbox{$\left#1\vbox to\h@Bigg{}\right.\n@space$}}}
%
%
%
\normalbaselineskip = 20pt plus 0.2pt minus 0.1pt
\normallineskip = 1.5pt plus 0.1pt minus 0.1pt
\normallineskiplimit = 1.5pt
\newskip\normaldisplayskip
\normaldisplayskip = 18pt plus 4pt minus 8pt
\newskip\normaldispshortskip
\normaldispshortskip = 5pt plus 4pt
\newskip\normalparskip
\normalparskip = 6pt plus 2pt minus 1pt
\newskip\skipregister
\skipregister = 5pt plus 2pt minus 1.5pt
\newif\ifsingl@    \newif\ifdoubl@
\newif\iftwelv@    \twelv@true
\def\singlespace{\singl@true\doubl@false\spaces@t}
\def\doublespace{\singl@false\doubl@true\spaces@t}
\def\normalspace{\singl@false\doubl@false\spaces@t}
\def\Tenpoint{\tenpoint\twelv@false\spaces@t}
\def\Twelvepoint{\twelvepoint\twelv@true\spaces@t}
\def\spaces@t{\relax%
 \iftwelv@ \ifsingl@\subspaces@t3:4;\else\subspaces@t1:1;\fi%
 \else \ifsingl@\subspaces@t3:5;\else\subspaces@t4:5;\fi \fi%
 \ifdoubl@ \multiply\baselineskip by 5%
 \divide\baselineskip by 4 \fi \unskip}
\def\subspaces@t#1:#2;{%
      \baselineskip = \normalbaselineskip%
      \multiply\baselineskip by #1 \divide\baselineskip by #2%
      \lineskip = \normallineskip%
      \multiply\lineskip by #1 \divide\lineskip by #2%
      \lineskiplimit = \normallineskiplimit%
      \multiply\lineskiplimit by #1 \divide\lineskiplimit by #2%
      \parskip = \normalparskip%
      \multiply\parskip by #1 \divide\parskip by #2%
      \abovedisplayskip = \normaldisplayskip%
      \multiply\abovedisplayskip by #1 \divide\abovedisplayskip by #2%
      \belowdisplayskip = \abovedisplayskip%
      \abovedisplayshortskip = \normaldispshortskip%
      \multiply\abovedisplayshortskip by #1%
        \divide\abovedisplayshortskip by #2%
      \belowdisplayshortskip = \abovedisplayshortskip%
      \advance\belowdisplayshortskip by \belowdisplayskip%
      \divide\belowdisplayshortskip by 2%
      \smallskipamount = \skipregister%
      \multiply\smallskipamount by #1 \divide\smallskipamount by #2%
      \medskipamount = \smallskipamount \multiply\medskipamount by 2%
      \bigskipamount = \smallskipamount \multiply\bigskipamount by 4 }
\def\normalbaselines{ \baselineskip=\normalbaselineskip%
   \lineskip=\normallineskip \lineskiplimit=\normallineskip%
   \iftwelv@\else \multiply\baselineskip by 4 \divide\baselineskip by 5%
     \multiply\lineskiplimit by 4 \divide\lineskiplimit by 5%
     \multiply\lineskip by 4 \divide\lineskip by 5 \fi }
\Twelvepoint  
\interlinepenalty=50
\interfootnotelinepenalty=5000
\predisplaypenalty=9000
\postdisplaypenalty=500
\hfuzz=1pt
\vfuzz=0.2pt
%
%
%
\def\pagecontents{%
   \ifvoid\topins\else\unvbox\topins\vskip\skip\topins\fi
   \dimen@ = \dp255 \unvbox255
   \ifvoid\footins\else\vskip\skip\footins\footrule\unvbox\footins\fi
   \ifr@ggedbottom \kern-\dimen@ \vfil \fi }
\def\makeheadline{\vbox to 0pt{ \skip@=\topskip
      \advance\skip@ by -12pt \advance\skip@ by -2\normalbaselineskip
      \vskip\skip@ \line{\vbox to 12pt{}\the\headline} \vss
      }\nointerlineskip}
\def\makefootline{\baselineskip = 1.5\normalbaselineskip
                 \line{\the\footline}}
\newif\iffrontpage
\newif\ifletterstyle
\newif\ifp@genum
\def\nopagenumbers{\p@genumfalse}
\def\pagenumbers{\p@genumtrue}
\pagenumbers
\newtoks\paperheadline
\newtoks\letterheadline
\newtoks\letterfrontheadline
\newtoks\lettermainheadline
\newtoks\paperfootline
\newtoks\letterfootline
\newtoks\date
\footline={\ifletterstyle\the\letterfootline\else\the\paperfootline\fi}
\paperfootline={\hss\iffrontpage\else\ifp@genum\tenrm\folio\hss\fi\fi}
\letterfootline={\hfil}
\headline={\ifletterstyle\the\letterheadline\else\the\paperheadline\fi}
\paperheadline={\hfil}
\letterheadline{\iffrontpage\the\letterfrontheadline
     \else\the\lettermainheadline\fi}
\lettermainheadline={\rm\ifp@genum page \ \folio\fi\hfil\the\date}
\def\monthname{\relax\ifcase\month 0/\or January\or February\or
   March\or April\or May\or June\or July\or August\or September\or
   October\or November\or December\else\number\month/\fi}
\date={\monthname\ \number\day, \number\year}
\countdef\pagenumber=1  \pagenumber=1
\def\advancepageno{\global\advance\pageno by 1
   \ifnum\pagenumber<0 \global\advance\pagenumber by -1
    \else\global\advance\pagenumber by 1 \fi \global\frontpagefalse }
\def\folio{\ifnum\pagenumber<0 \romannumeral-\pagenumber
           \else \number\pagenumber \fi }
\def\footrule{\dimen@=\prevdepth\nointerlineskip
   \vbox to 0pt{\vskip -0.25\baselineskip \hrule width 0.35\hsize \vss}
   \prevdepth=\dimen@ }
\newtoks\foottokens
\foottokens={\Tenpoint\singlespace}
\newdimen\footindent
\footindent=24pt
\def\vfootnote#1{\insert\footins\bgroup  \the\foottokens
   \interlinepenalty=\interfootnotelinepenalty \floatingpenalty=20000
   \splittopskip=\ht\strutbox \boxmaxdepth=\dp\strutbox
   \leftskip=\footindent \rightskip=\z@skip
   \parindent=0.5\footindent \parfillskip=0pt plus 1fil
   \spaceskip=\z@skip \xspaceskip=\z@skip
   \Textindent{$ #1 $}\footstrut\futurelet\next\fo@t}
\def\Textindent#1{\noindent\llap{#1\enspace}\ignorespaces}
\def\footnote#1{\attach{#1}\vfootnote{#1}}

\let\footsymbol=\star
\newcount\lastf@@t           \lastf@@t=-1
\newcount\footsymbolcount    \footsymbolcount=0
\newif\ifPhysRev
\def\footsymbolgen{\relax \ifPhysRev \iffrontpage \NPsymbolgen\else
      \PRsymbolgen\fi \else \NPsymbolgen\fi
   \global\lastf@@t=\pageno \footsymbol }
\def\NPsymbolgen{\ifnum\footsymbolcount<0 \global\footsymbolcount=0\fi
   {\iffrontpage \else \advance\lastf@@t by 1 \fi
    \ifnum\lastf@@t<\pageno \global\footsymbolcount=0
     \else \global\advance\footsymbolcount by 1 \fi }
   \ifcase\footsymbolcount \fd@f\star\or \fd@f\dagger\or \fd@f\ddagger\or
    \fd@f\ast\or \fd@f\natural\or \fd@f\diamond\or \fd@f\bullet\or
    \fd@f\nabla\else \fd@f\dagger\global\footsymbolcount=0 \fi }
\def\fd@f#1{\xdef\footsymbol{#1}}
\def\PRsymbolgen{\ifnum\footsymbolcount>0 \global\footsymbolcount=0\fi
      \global\advance\footsymbolcount by -1
      \xdef\footsymbol{\sharp\number-\footsymbolcount} }
\def\space@ver#1{\let\@sf=\empty \ifmmode #1\else \ifhmode
   \edef\@sf{\spacefactor=\the\spacefactor}\unskip${}#1$\relax\fi\fi}
\def\attach#1{\space@ver{\strut^{\mkern 2mu #1} }\@sf\ }
%
%
%
\newcount\chapternumber      \chapternumber=0
\newcount\sectionnumber      \sectionnumber=0
\newcount\equanumber         \equanumber=0
\let\chapterlabel=0
\newtoks\chapterstyle        \chapterstyle={\Number}
\newskip\chapterskip         \chapterskip=\bigskipamount
\newskip\sectionskip         \sectionskip=\medskipamount
\newskip\headskip            \headskip=8pt plus 3pt minus 3pt
\newdimen\chapterminspace    \chapterminspace=15pc
\newdimen\sectionminspace    \sectionminspace=10pc
\newdimen\referenceminspace  \referenceminspace=25pc
\def\chapterreset{\global\advance\chapternumber by 1
   \ifnum\equanumber<0 \else\global\equanumber=0\fi
   \sectionnumber=0 \makel@bel}
\def\makel@bel{\xdef\chapterlabel{%
\the\chapterstyle{\the\chapternumber}.}}
\def\sectionlabel{\number\sectionnumber \quad }
\def\alphabetic#1{\count255='140 \advance\count255 by #1\char\count255}
\def\Alphabetic#1{\count255='100 \advance\count255 by #1\char\count255}
\def\Roman#1{\uppercase\expandafter{\romannumeral #1}}
\def\roman#1{\romannumeral #1}
\def\Number#1{\number #1}
\def\unnumberedchapters{\let\makel@bel=\relax \let\chapterlabel=\relax
\let\sectionlabel=\relax \equanumber=-1 }
\def\titlestyle#1{\par\begingroup \interlinepenalty=9999
     \leftskip=0.02\hsize plus 0.23\hsize minus 0.02\hsize
     \rightskip=\leftskip \parfillskip=0pt
     \hyphenpenalty=9000 \exhyphenpenalty=9000
     \tolerance=9999 \pretolerance=9000
     \spaceskip=0.333em \xspaceskip=0.5em
     \iftwelv@\twelvepoint\fourteenrm\else\twelvepoint\fi
   \noindent #1\par\endgroup }
\def\spacecheck#1{\dimen@=\pagegoal\advance\dimen@ by -\pagetotal
   \ifdim\dimen@<#1 \ifdim\dimen@>0pt \vfil\break \fi\fi}
\def\chapter#1{\par \penalty-300 \vskip\chapterskip
   \spacecheck\chapterminspace
   \chapterreset \titlestyle{\chapterlabel \ #1}
   \nobreak\vskip\headskip \penalty 30000
   \wlog{\string\chapter\ \chapterlabel} }

\def\section#1{\par \ifnum\the\lastpenalty=30000\else
   \penalty-200\vskip\sectionskip \spacecheck\sectionminspace\fi
   \wlog{\string\section\ \chapterlabel \the\sectionnumber}
   \global\advance\sectionnumber by 1  \noindent
   {\caps\enspace\chapterlabel \sectionlabel #1}\par
   \nobreak\vskip\headskip \penalty 30000 }
\def\subsection#1{\par
   \ifnum\the\lastpenalty=30000\else \penalty-100\smallskip \fi
   \noindent\undertext{#1}\enspace \vadjust{\penalty5000}}

\def\undertext#1{\vtop{\hbox{#1}\kern 1pt \hrule}}
\def\APPENDIX#1#2{\par\penalty-300\vskip\chapterskip
   \spacecheck\chapterminspace \chapterreset \xdef\chapterlabel{#1}
   \titlestyle{APPENDIX #2} \nobreak\vskip\headskip \penalty 30000
   \wlog{\string\Appendix\ \chapterlabel} }
\def\Appendix#1{\APPENDIX{#1}{#1}}
\def\appendix{\APPENDIX{A}{}}
%
%
%

\newif\ifdraftmode
\draftmodefalse

\def\eqname#1{\relax \ifnum\equanumber<0
     \xdef#1{{(\number-\equanumber)}}\global\advance\equanumber by -1
    \else \global\advance\equanumber by 1
      \xdef#1{{(\chapterlabel \number\equanumber)}} \fi}
%



%
\def\eqinsert#1{\noalign{\dimen@=\prevdepth \nointerlineskip
   \setbox0=\hbox to\displaywidth{\hfil #1}
   \vbox to 0pt{\vss\hbox{$\!\box0\!$}\kern-0.5\baselineskip}
   \prevdepth=\dimen@}}
\def\sequentialequations{\equanumber=-1}
%
%
\def\GENITEM#1;#2{\par \hangafter=0 \hangindent=#1
    \Textindent{$ #2 $}\ignorespaces}
\outer\def\newitem#1=#2;{\gdef#1{\GENITEM #2;}}
\newdimen\itemsize                \itemsize=30pt
\newitem\item=1\itemsize;
\newitem\sitem=1.75\itemsize;     
\newitem\ssitem=2.5\itemsize;     
\outer\def\newlist#1=#2&#3&#4;{\toks0={#2}\toks1={#3}%
   \count255=\escapechar \escapechar=-1
   \alloc@0\list\countdef\insc@unt\listcount     \listcount=0
   \edef#1{\par
      \countdef\listcount=\the\allocationnumber
      \advance\listcount by 1
      \hangafter=0 \hangindent=#4
      \Textindent{\the\toks0{\listcount}\the\toks1}}
   \expandafter\expandafter\expandafter
    \edef\c@t#1{begin}{\par
      \countdef\listcount=\the\allocationnumber \listcount=1
      \hangafter=0 \hangindent=#4
      \Textindent{\the\toks0{\listcount}\the\toks1}}
   \expandafter\expandafter\expandafter
    \edef\c@t#1{con}{\par \hangafter=0 \hangindent=#4 \noindent}
   \escapechar=\count255}
\def\c@t#1#2{\csname\string#1#2\endcsname}
\newlist\point=\Number&.&1.0\itemsize;
\newlist\subpoint=(\alphabetic&)&1.75\itemsize;
\newlist\subsubpoint=(\roman&)&2.5\itemsize;
\let\spoint=\subpoint

%
%
%
\newcount\referencecount     \referencecount=0
\newif\ifreferenceopen       \newwrite\referencewrite
\newtoks\rw@toks
\def\NPrefmark#1{\attach{\scriptscriptstyle [ #1 ] }}
\let\PRrefmark=\attach
\def\refmark#1{\relax\ifPhysRev\PRrefmark{#1}\else\NPrefmark{#1}\fi}
\def\refend{\refmark{\number\referencecount}}
\newcount\lastrefsbegincount \lastrefsbegincount=0
\def\refsend{\refmark{\count255=\referencecount
   \advance\count255 by-\lastrefsbegincount
   \ifcase\count255 \number\referencecount
   \or \number\lastrefsbegincount,\number\referencecount
   \else \number\lastrefsbegincount-\number\referencecount \fi}}
\def\refch@ck{\chardef\rw@write=\referencewrite
   \ifreferenceopen \else \referenceopentrue
   \immediate\openout\referencewrite=referenc.tex \fi}
%
{\catcode`\^^M=\active 
  \gdef\obeyendofline{\catcode`\^^M\active \let^^M\ }}%
%
{\catcode`\^^M=\active 
  \gdef\ignoreendofline{\catcode`\^^M=5}}
{\obeyendofline\gdef\rw@start#1{\def\t@st{#1} \ifx\t@st\blankend%
\endgroup \@sf \relax \else \ifx\t@st\bl@nkend \endgroup \@sf \relax%
\else \rw@begin#1
\backtotext
\fi \fi } }
{\obeyendofline\gdef\rw@begin#1
{\def\n@xt{#1}\rw@toks={#1}\relax%
\rw@next}}
\def\blankend{}
{\obeylines\gdef\bl@nkend{
}}
\newif\iffirstrefline  \firstreflinetrue
\def\rwr@teswitch{\ifx\n@xt\blankend \let\n@xt=\rw@begin %
 \else\iffirstrefline \global\firstreflinefalse%
\immediate\write\rw@write{\noexpand\obeyendofline \the\rw@toks}%
\let\n@xt=\rw@begin%
      \else\ifx\n@xt\rw@@d \def\n@xt{\immediate\write\rw@write{%
        \noexpand\ignoreendofline}\endgroup \@sf}%
             \else \immediate\write\rw@write{\the\rw@toks}%
             \let\n@xt=\rw@begin\fi\fi \fi}
\def\rw@next{\rwr@teswitch\n@xt}
\def\rw@@d{\backtotext} \let\rw@end=\relax
\let\backtotext=\relax

\newdimen\refindent     \refindent=30pt
\def\refitem#1{\par \hangafter=0 \hangindent=\refindent \Textindent{#1}}
\def\REFNUM#1{\space@ver{}\refch@ck \firstreflinetrue%
 \global\advance\referencecount by 1 \xdef#1{\the\referencecount}}
\def\refnum#1{\space@ver{}\refch@ck \firstreflinetrue%
 \global\advance\referencecount by 1 \xdef#1{\the\referencecount}\refend}

\def\REF#1{\REFNUM#1%
 \immediate\write\referencewrite{%
            \noexpand\refitem{\ifdraftmode{\sevenrm
               \noexpand\string\string#1\ }\fi#1.}}%
      \begingroup\obeyendofline\rw@start}
\def\ref{\refnum\?%
 \immediate\write\referencewrite{\noexpand\refitem{\?.}}%
\begingroup\obeyendofline\rw@start}
\def\Ref#1{\refnum#1%
 \immediate\write\referencewrite{\noexpand\refitem{#1.}}%
\begingroup\obeyendofline\rw@start}
\def\REFS#1{\REFNUM#1\global\lastrefsbegincount=\referencecount
\immediate\write\referencewrite{\noexpand\refitem{#1.}}%
\begingroup\obeyendofline\rw@start}
%

%
%

%
\newcount\figurecount     \figurecount=0
\newif\iffigureopen       \newwrite\figurewrite
\def\figch@ck{\chardef\rw@write=\figurewrite \iffigureopen\else
   \immediate\openout\figurewrite=figures.aux
   \figureopentrue\fi}
\def\FIGNUM#1{\space@ver{}\figch@ck \firstreflinetrue%
 \global\advance\figurecount by 1 \xdef#1{\the\figurecount}}
\def\FIG#1{\FIGNUM#1
   \immediate\write\figurewrite{\noexpand\refitem{#1.}}%
   \begingroup\obeyendofline\rw@start}
\def\fig{\FIGNUM\? fig.~\?%
\immediate\write\figurewrite{\noexpand\refitem{\?.}}%
\begingroup\obeyendofline\rw@start}
\def\figure{\FIGNUM\? figure~\?
   \immediate\write\figurewrite{\noexpand\refitem{\?.}}%
   \begingroup\obeyendofline\rw@start}
\def\Fig{\FIGNUM\? Fig.~\?%
\immediate\write\figurewrite{\noexpand\refitem{\?.}}%
\begingroup\obeyendofline\rw@start}
\def\Figure{\FIGNUM\? Figure~\?%
\immediate\write\figurewrite{\noexpand\refitem{\?.}}%
\begingroup\obeyendofline\rw@start}
\newcount\tablecount     \tablecount=0
\newif\iftableopen       \newwrite\tablewrite
\def\tabch@ck{\chardef\rw@write=\tablewrite \iftableopen\else
   \immediate\openout\tablewrite=tables.aux
   \tableopentrue\fi}
\def\TABNUM#1{\space@ver{}\tabch@ck \firstreflinetrue%
 \global\advance\tablecount by 1 \xdef#1{\the\tablecount}}
\def\TABLE#1{\TABNUM#1
   \immediate\write\tablewrite{\noexpand\refitem{#1.}}%
   \begingroup\obeyendofline\rw@start}
\def\Table{\TABNUM\? Table~\?%
\immediate\write\tablewrite{\noexpand\refitem{\?.}}%
\begingroup\obeyendofline\rw@start}
\newif\ifsymbolopen       \newwrite\symbolwrite
\def\symch@ck{\ifsymbolopen\else
   \immediate\openout\symbolwrite=symbols.aux
   \symbolopentrue\fi}
\def\symdef#1#2{\def#1{#2}%
      \symch@ck%
      \immediate\write\symbolwrite{$$ \hbox{\noexpand\string\string#1}
                \noexpand\qquad\noexpand\longrightarrow\noexpand\qquad
                           \string#1 $$}}
%
%

%
%
%
\def\masterreset{\global\pagenumber=1 \global\chapternumber=0
   \global\equanumber=0 \global\sectionnumber=0
   \global\referencecount=0 \global\figurecount=0 \global\tablecount=0 }
\def\FRONTPAGE{\ifvoid255\else\vfill\penalty-2000\fi
      \masterreset\global\frontpagetrue
      \global\lastf@@t=0 \global\footsymbolcount=0}

\def\paperstyle{\letterstylefalse\normalspace\papersize}
\def\letterstyle{\letterstyletrue\singlespace\lettersize}
\def\papersize{\hsize=35pc\vsize=50pc\hoffset=1pc\voffset=6pc
               \skip\footins=\bigskipamount}
\def\lettersize{\hsize=6.5in\vsize=8.5in\hoffset=0in\voffset=1in
   \skip\footins=\smallskipamount \multiply\skip\footins by 3 }
\paperstyle   
%
%
\def\MEMO{\letterstyle\FRONTPAGE \letterfrontheadline={\hfil}
    \line{\quad\fourteenrm NTC MEMORANDUM\hfil\twelverm\the\date\quad}
    \medskip \memod@f}

\def\memit@m#1{\smallskip \hangafter=0 \hangindent=1in
      \Textindent{\caps #1}}
\def\memod@f{\xdef\to{\memit@m{To:}}\xdef\from{\memit@m{From:}}%
     \xdef\topic{\memit@m{Topic:}}\xdef\subject{\memit@m{Subject:}}%
     \xdef\rule{\bigskip\hrule height 1pt\bigskip}}
\memod@f
\newskip\lettertopfil
\lettertopfil = 0pt plus 1.5in minus 0pt
\newskip\letterbottomfil
\letterbottomfil = 0pt plus 2.3in minus 0pt
\newskip\spskip \setbox0\hbox{\ } \spskip=-1\wd0
\def\addressee#1{\medskip\rightline{\the\date\hskip 30pt} \bigskip
   \vskip\lettertopfil
   \ialign to\hsize{\strut ##\hfil\tabskip 0pt plus \hsize \cr #1\crcr}
   \medskip\noindent\hskip\spskip}
\newskip\signatureskip       \signatureskip=40pt
\def\signed#1{\par \penalty 9000 \bigskip \dt@pfalse
  \everycr={\noalign{\ifdt@p\vskip\signatureskip\global\dt@pfalse\fi}}
  \setbox0=\vbox{\singlespace \halign{\tabskip 0pt \strut ##\hfil\cr
   \noalign{\global\dt@ptrue}#1\crcr}}
  \line{\hskip 0.5\hsize minus 0.5\hsize \box0\hfil} \medskip }

\def\endletter{\ifnum\pagenumber=1 \vskip\letterbottomfil\supereject
\else \vfil\supereject \fi}
\newbox\letterb@x
\def\lettertext{\par\unvcopy\letterb@x\par}
\def\multiletter{\setbox\letterb@x=\vbox\bgroup
      \everypar{\vrule height 1\baselineskip depth 0pt width 0pt }
      \singlespace \topskip=\baselineskip }
\def\letterend{\par\egroup}
%
%
%
\newskip\frontpageskip
\newtoks\pubtype
\newtoks\Pubnum
\newtoks\pubnum
\newif\ifp@bblock  \p@bblocktrue
\def\PH@SR@V{\doubl@true \baselineskip=24.1pt plus 0.2pt minus 0.1pt
             \parskip= 3pt plus 2pt minus 1pt }
\def\PHYSREV{\paperstyle\PhysRevtrue\PH@SR@V}
\def\titlepage{\FRONTPAGE\paperstyle\ifPhysRev\PH@SR@V\fi
   \ifp@bblock\p@bblock\fi}
\def\nopubblock{\p@bblockfalse}
\def\endpage{\vfil\break}
\frontpageskip=1\medskipamount plus .5fil
\pubtype={\tensl Preliminary Version}
%
\def\p@bblock{\begingroup \tabskip=\hsize minus \hsize
   \baselineskip=1.5\ht\strutbox \topspace-2\baselineskip
   \halign to\hsize{\strut ##\hfil\tabskip=0pt\crcr
   \the\Pubnum\cr \the\date\cr \the\pubtype\cr}\endgroup}
\def\title#1{\vskip\frontpageskip \titlestyle{#1} \vskip\headskip }
\def\author#1{\vskip\frontpageskip\titlestyle{\twelvecp #1}\nobreak}

\def\address#1{\par\kern 5pt\titlestyle{\twelvepoint\it #1}}
\def\andaddress{\par\kern 5pt \centerline{\sl and} \address}

\def\abstract{\vskip\frontpageskip\centerline{\fourteenrm ABSTRACT}
              \vskip\headskip }

%
%
%

\def\\{\relax\ifmmode\backslash\else$\backslash$\fi}
\def\globaleqnumbers{\relax\if\equanumber<0\else\global\equanumber=-1\fi}

\def\journal#1&#2(#3){\unskip, \sl #1~\bf #2 \rm (19#3) }

\def\topspace{\hrule height 0pt depth 0pt \vskip}

\let\int=\intop         
\def\prop{\mathrel{{\mathchoice{\pr@p\scriptstyle}{\pr@p\scriptstyle}{
                \pr@p\scriptscriptstyle}{\pr@p\scriptscriptstyle} }}}
\def\pr@p#1{\setbox0=\hbox{$\cal #1 \char'103$}
   \hbox{$\cal #1 \char'117$\kern-.4\wd0\box0}}
\def\lsim{\mathrel{\mathpalette\@versim<}}
\def\gsim{\mathrel{\mathpalette\@versim>}}
\def\@versim#1#2{\lower0.5ex\vbox{\baselineskip\z@skip\lineskip-.1ex
  \lineskiplimit\z@\ialign{$\m@th#1\hfil##\hfil$\crcr#2\crcr\sim\crcr}}}
%
%
%
\let\sec@nt=\sec
\def\sec{\relax\ifmmode\let\n@xt=\sec@nt\else\let\n@xt\section\fi\n@xt}
\def\obsolete#1{\message{Macro \string #1 is obsolete.}}
\def\firstsec#1{\obsolete\firstsec \section{#1}}
\def\firstsubsec#1{\obsolete\firstsubsec \subsection{#1}}
\def\thispage#1{\obsolete\thispage \global\pagenumber=#1\frontpagefalse}
\def\thischapter#1{\obsolete\thischapter \global\chapternumber=#1}
\def\nextequation#1{\obsolete\nextequation \global\equanumber=#1
   \ifnum\the\equanumber>0 \global\advance\equanumber by 1 \fi}
\def\BOXITEM{\afterassigment\B@XITEM\setbox0=}
\def\B@XITEM{\par\hangindent\wd0 \noindent\box0 }
%

%
%
%
%
%
\lock
\message{   }
%
%
%












\newbox\figbox
\newdimen\zero  \zero=0pt
\newdimen\figmove
\newdimen\figwidth
\newdimen\figheight
\newdimen\textwidth
\newtoks\figtoks
\newcount\figcounta
\newcount\figcountb
\newcount\figlines
\def\figreset{\global\figcounta=-1 \global\figcountb=-1
\global\figmove=\baselineskip
\global\figlines=1 \global\figtoks={ } }
\def\picture#1by#2:#3{\global\setbox\figbox=\vbox{\vskip #1
\hbox{\vbox{\hsize=#2 \noindent #3}}}
\global\setbox\figbox=\vbox{\kern 10pt
\hbox{\kern 10pt \box\figbox \kern 10pt }\kern 10pt}
\global\figwidth=1\wd\figbox
\global\figheight=1\ht\figbox
\global\textwidth=\hsize
\global\advance\textwidth by - \figwidth }
\def\figtoksappend{\edef\temp##1{\global\figtoks=%
{\the\figtoks ##1}}\temp}
\def\figparmsa#1{\loop \global\advance\figcounta by 1
\ifnum \figcounta < #1
\figtoksappend{ 0pt \the\hsize }
\global\advance\figlines by 1
\repeat }
\def\figparmsb#1{\loop \global\advance\figcountb by 1
\ifnum \figcountb < #1
\figtoksappend{ \the\figwidth \the\textwidth}
\global\advance\figlines by 1
\repeat }
\def\figtext#1:#2:#3{\figreset%
\figparmsa{#1}%
\figparmsb{#2}%
\multiply\figmove by #1%
\global\setbox\figbox=\vbox to 0pt{\vskip \figmove  \hbox{\box\figbox}
\vss }
\parshape=\the\figlines\the\figtoks\the\zero\the\hsize
\noindent
\rlap{\box\figbox} #3}
\def\Buildrel#1\under#2{\mathrel{\mathop{#2}\limits_{#1}}}
\def\llongrarrow{\hbox to 40pt{\rightarrowfill}}



\def\boxit#1{\vbox{\hrule\hbox{\vrule\kern3pt
\vbox{\kern3pt#1\kern3pt}\kern3pt\vrule}\hrule}}
\newdimen\str
\def\fboxit#1#2{\vbox{\hrule height #1 \hbox{\vrule width #1
\kern3pt \vbox{\kern3pt#2\kern3pt}\kern3pt \vrule width #1 }
\hrule height #1 }}
\def\tran#1#2{\transpoint \hfuzz 5pt \gdef\label{#1}
\vbox to \the\vsize{\hsize \the\hsize #2} \par \eject }
\newdimen\baseskip
\newdimen\lskip
\lskip=\lineskip
\newdimen\transize
\newdimen\tall
\def\transpoint{\gdef\rm{\fam0\eighteenrm}
\font\twentyfourit = cmti10 scaled \magstep5
\font\twentyfourrm = cmr10 scaled \magstep5
\font\twentyfourbf = cmbx10 scaled \magstep5
\font\twentyeightsy = cmsy10 scaled \magstep5
\font\eighteenrm = cmr10 scaled \magstep3
\font\eighteenb = cmbx10 scaled \magstep3
\font\eighteeni = cmmi10 scaled \magstep3
\font\eighteenit = cmti10 scaled \magstep3
\font\eighteensl = cmsl10 scaled \magstep3
\font\eighteensy = cmsy10 scaled \magstep3
\font\eighteencaps = cmr10 scaled \magstep3
\font\eighteenmathex = cmex10 scaled \magstep3
\font\fourteenrm=cmr10 scaled \magstep2
\font\fourteeni=cmmi10 scaled \magstep2
\font\fourteenit = cmti10 scaled \magstep2
\font\fourteensy=cmsy10 scaled \magstep2
\font\fourteenmathex = cmex10 scaled \magstep2
\parindent 20pt
\global\hsize = 7.0in
\global\vsize = 8.9in
\dimen\transize = \the\hsize
\dimen\tall = \the\vsize
\def\sy{\eighteensy }
\def\sl{\eighteens }
\def\bf{\eighteenb }
\def\it{\eighteenit }
\def\caps{\eighteencaps }
\textfont0=\eighteenrm \scriptfont0=\fourteenrm
\scriptscriptfont0=\twelverm
\textfont1=\eighteeni \scriptfont1=\fourteeni \scriptscriptfont1=\twelvei
\textfont2=\eighteensy \scriptfont2=\fourteensy
\scriptscriptfont2=\twelvesy
\textfont3=\eighteenmathex \scriptfont3=\eighteenmathex
\scriptscriptfont3=\eighteenmathex
\global\baselineskip 35pt
\global\lineskip 15pt
\global\parskip 5pt  plus 1pt minus 1pt
\global\abovedisplayskip  3pt plus 10pt minus 10pt
\global\belowdisplayskip 3pt plus 10pt minus 10pt
\def\rtitle##1{\centerline{\undertext{\twentyfourrm ##1}}}
\def\ititle##1{\centerline{\undertext{\twentyfourit ##1}}}
\def\ctitle##1{\centerline{\undertext{\caps ##1}}}
\def\vstrut{\hbox{\vrule width 0pt height .35in depth .15in }}
\def\cline##1{\centerline{\vstrut ##1}}
\output{\shipout\vbox{\vskip .5in
\pagecontents \vfill
\hbox to \the\hsize{\hfill{\tenbf \label} } }
\global\advance\count0 by 1 }
\rm }


%
%
%

%

%

%

%
{\obeyspaces\global\let =\ }
%
%
%
\widowpenalty 1000
\thickmuskip 4mu plus 4mu
\unlock
%
%
\def\p@nnlock{\begingroup \tabskip=\hsize minus \hsize
   \baselineskip=1.5\ht\strutbox \topspace-2\baselineskip
   \noindent\strut\the\Pubnum \hfill \the\date   \endgroup}
\def\titlepage{\FRONTPAGE\paperstyle\p@nnlock}
\def\displaylines#1{\displ@y
  \halign{\hbox to\displaywidth{$\hfil\displaystyle##\hfil$}\crcr
    #1\crcr}}
\def\addressee#1{\null
   \bigskip\medskip\rightline{\the\date\hskip 30pt}
   \vskip\lettertopfil
   \ialign to\hsize{\strut ##\hfil\tabskip 0pt plus \hsize \cr #1\crcr}
   \medskip\vskip 3pt\noindent}
\def\tmsaddressee#1#2{
   \vskip\lettertopfil
  \setbox0=\vbox{\singlespace \halign{\tabskip 0pt \strut ##\hfil\cr
   \noalign{\global\dt@ptrue}#1\crcr}}
  \line{\hskip 0.7\hsize minus 0.7\hsize \box0\hfil}
   \bigskip
   \vskip .2in
   \ialign to\hsize{\strut ##\hfil\tabskip 0pt plus \hsize \cr #2\crcr}
   \medskip\vskip 3pt\noindent}
\def\makeheadline{\vbox to 0pt{ \skip@=\topskip
      \advance\skip@ by -12pt \advance\skip@ by -2\normalbaselineskip
      \vskip\skip@  \vss
      }\nointerlineskip}
\def\signed#1{\par \penalty 9000 \bigskip \vskip .06in\dt@pfalse
  \everycr={\noalign{\ifdt@p\vskip\signatureskip\global\dt@pfalse\fi}}
  \setbox0=\vbox{\singlespace \halign{\tabskip 0pt \strut ##\hfil\cr
   \noalign{\global\dt@ptrue}#1\crcr}}
  \line{\hskip 0.5\hsize minus 0.5\hsize \box0\hfil} \medskip }
\def\lettersize{\hsize=6.25in\vsize=8.5in\hoffset=0in\voffset=1in
   \skip\footins=\smallskipamount \multiply\skip\footins by 3 }
%
%
%
%
%
\outer\def\newnewlist#1=#2&#3&#4&#5;{\toks0={#2}\toks1={#3}%
   \dimen1=\hsize  \advance\dimen1 by -#4
   \dimen2=\hsize  \advance\dimen2 by -#5
   \count255=\escapechar \escapechar=-1
   \alloc@0\list\countdef\insc@unt\listcount     \listcount=0
   \edef#1{\par
      \countdef\listcount=\the\allocationnumber
      \advance\listcount by 1
      \parshape=2 #4 \dimen1 #5 \dimen2
      \Textindent{\the\toks0{\listcount}\the\toks1}}
   \expandafter\expandafter\expandafter
    \edef\c@t#1{begin}{\par
      \countdef\listcount=\the\allocationnumber \listcount=1
      \parshape=2 #4 \dimen1 #5 \dimen2
      \Textindent{\the\toks0{\listcount}\the\toks1}}
   \expandafter\expandafter\expandafter
    \edef\c@t#1{con}{\par \parshape=2 #4 \dimen1 #5 \dimen2 \noindent}
   \escapechar=\count255}
\def\c@t#1#2{\csname\string#1#2\endcsname}
%
%
%
%
%
%
%
\def\noparGENITEM#1;{\hangafter=0 \hangindent=#1
    \ignorespaces\noindent}
\outer\def\noparnewitem#1=#2;{\gdef#1{\noparGENITEM #2;}}
\noparnewitem\spoint=1.5\itemsize;
%
%
%
\def\MEMO{\letterstyle\FRONTPAGE \letterfrontheadline={\hfil}
      \hoffset=1in \voffset=1.21in
    \line{\hskip .8in  \special{overlay ntcmemo.dat}
          \quad\fourteenrm NTC MEMORANDUM\hfil\twelverm\the\date\quad}
    \medskip\medskip \memod@f}

\def\memit@m#1{\smallskip \hangafter=0 \hangindent=1in
      \Textindent{\caps #1}}
\def\memod@f{\xdef\to{\memit@m{To:}}\xdef\from{\memit@m{From:}}%
     \xdef\topic{\memit@m{Topic:}}\xdef\subject{\memit@m{Subject:}}%
     \xdef\rule{\bigskip\hrule height 1pt\bigskip}}
\memod@f
\lock
\def\NPrefmark#1{\attach{\scriptscriptstyle  #1 }}

%

%
%
%
\def\papersize{\hsize=6.5in\vsize=8.60in\hoffset=0.2in\voffset=2pc
                \skip\footins=\bigskipamount}
\paperstyle
%
%
\doublespace
\pretolerance=500
\tolerance=500
\widowpenalty 2000
\thinmuskip=4mu
\medmuskip=5mu plus 2mu minus 4mu
\thickmuskip=7mu plus 5mu
\PHYSREV
%
%
\REF\EMCX{J. J. Aubert et al. (EMC collaboration),
                          Phys. Lett. {\bf 123B}, 275 (1983).}
\REF\RESCALE{F. E. Close, R. G. Roberts, and G. G. Ross,
                          Phys. Lett. {\bf 129B}, 346 (1983);
                          Nucl. Phys. {\bf B296}, 582 (1988);
             F. E. Close, R. L. Jaffe, R. G. Roberts, and G. G. Ross,
                          Phys. Rev. {\bf D31}, 1004 (1985). }
\REF\ARGONNE{E. L. Berger and F. Coester,
                        Annu. Rev. Nucl. Part. Sci. {\bf 37}, 463 (1987);
             R. P. Bickerstaff and A. W. Thomas,
                        J. Phys. G {\bf 15}, 1523 (1989);
        R. P. Bickerstaff, M. C. Birse, and G. A. Miller,
                      Phys. Rev. {\bf D33}, 3228 (1986).}
\REF\FNAL{D. M. Alde et al. (E772 collaboration),
                            Phys. Rev. Lett. {\bf 64}, 2479 (1990);
          A. Magnon, talk given at the workshop on
                 Baryon Spectroscopy and the Structure of the Nucleon,
                 Saclay, France, Sept. 23$-$25, 1991.}
\REF\SKFEC{S. Kumano and F. E. Close,
               Phys. Rev. {\bf C41}, 1855 (1990).}
\REF\SK{S. Kumano, Indiana University preprint IU/NTC-92-02;
                 talk given at the 13th International Conference
                            on Few Body Problems in Physics,
                            Adelaide, Australia,
                            Jan. 5$-$11, 1992;
                 in Proceedings of the International Workshop
                 on Gross Properties
                            of Nuclei and Nuclear Excitations,
                            Hirschegg, Austria, Jan. 20$-$25, 1992,
                            edited by H. Feldmeier;
                 to be submitted for publication.}
\REF\MONIZ{T. de Forest and P. J. Mulders,
                            Phys. Rev. {\bf D35}, 2849 (1987);
           S. Kumano and E. J. Moniz, Phys. Rev. {\bf C37}, 2088 (1988).}
\REF\NMC{P. Amaudruz et al. (NMC collaboration),
                            Z. Phys. {\bf C51}, 387 (1991);
         M. van der Heijden, Ph.D thesis, University of Amsterdam (1991);
         M. A. J. Botje and C. Scholz
                       in
      {\it Intersections between Particle and Nuclear Physics},
                            edited by W. T. H. van Oers,
                            American Institute of Physics (1991).}
\REF\ESIX{M. R. Adams et al. (E665 collaboration),
                           Phys. Rev. Lett. {\bf 68}, 3266 (1992);
          D. E. Jaffe, in
      {\it Intersections between Particle and Nuclear Physics},
                            edited by W. T. H. van Oers,
                            American Institute of Physics (1991);
          C. W. Salgado, talk given at the workshop on
                            High Energy Probes of QCD and Nuclei,
                            University Park, Pennsylvania, March 25-28, 1992.}
\REF\EMCA{M. Arneodo et al. (EMC collaboration),
                            Nucl. Phys. {\bf B333}, 1 (1990);
                            Phys. Lett. {\bf 211B}, 493 (1988). }
\REF\VECTOR{For recent investigations on the vector meson
                 dominance model, see
            C. L. Bilchak, D. Schildknecht, and J. D. Stroughair,
                         Phys. Lett. {\bf 214B}, 441 (1988);
                                     {\bf 233B}, 461 (1989);
            G. Shaw,     Phys. Lett. {\bf 228B}, 125 (1989);
            G. Piller and W. Weise,
                         Phys. Rev. {\bf C42}, R1834 (1990).}
\REF\RECOMB{N. N. Nicolaev and V. I. Zakharov,
                         Phys. Lett. {\bf 55B}, 397 (1975).}
\REF\MQ{ A. H. Mueller and J. Qiu, Nucl. Phys. {\bf B268}, 427 (1986);
                          J. Qiu, Nucl. Phys. {\bf B291}, 746 (1987);
            E. L. Berger and J. Qiu,
                         Phys. Lett. {\bf 206B}, 141 (1988).}
\REF\CR{    F. E. Close and R. G. Roberts,
                         Phys. Lett. {\bf 213B}, 91 (1988).}
\REF\PREDA{R. J. M. Covolan and E. Predazzi,
                         Nuovo Cimento, {\bf 103A}, 773 (1990);
            W. Zhu and J. G. Shen,
                         Phys. Lett. {\bf 235B}, 170 (1990);
                         J. Phys. {\bf G16}, 925 (1990);
            G. Li, Z. Cao, and C. Zhong,
                         Nucl. Phys. {\bf A509}, 757 (1990);
            M. Altmann, M. Gl\"uck, and E. Reya,
                         Phys. Lett. {\bf 285B}, 359 (1992).}
\REF\LIUTI{L. L. Frankfurt, S. Liuti, and M. I. Strikman,
           research in progress; S. Liuti, personal communication;
           V. Barone et al., preprint KFA-IKP(TH)-1992-13.}
\REF\CQR{F. E. Close, J. Qiu, and R. G. Roberts,
                          Phys. Rev. {\bf D40}, 2820 (1989). }
\REF\FF{ L. L. Frankfurt and M. I. Strikman,
                         Phys. Rep.  {\bf 160}, 235 (1988);
                         Nucl. Phys. {\bf B316}, 340 (1989);
                         Phys. Rev. Lett. {\bf 65}, 1725 (1990);
         S. J. Brodsky and H. J. Lu,
                         Phys. Rev. Lett. {\bf 64}, 1342 (1990);
         N. N. Nikolaev and B. G. Zakharov,
                         Phys. Lett. {\bf 260B}, 414 (1991);
                         Z. Phys. {\bf C49}, 607 (1991);
         V. R. Zoller,   Z. Phys. {\bf C53}, 443 (1992);
                         Phys. Lett. {\bf 279B}, 145 (1992).}
\REF\POMERON{P. Castorina and A. Donnachie,
                         Phys. Lett. {\bf 215B}, 589 (1988);
                         Z. Phys. {\bf C45}, 141 (1989);
          J. Kwiecinski, Z. Phys. {\bf C45}, 461 (1990).}
\REF\PRE{G. Preparata and P. G. Ratcliffe, preprint MITH91-13.}
\REF\MIX{J. Kwiecinski and B. Badelek,
                         Phys. Lett. {\bf 208B}, 508 (1988);
         B. Badelek and J. Kwiecinski, Nucl. Phys. {\bf B370}, 278 (1992);
         W. Melnitchouk and A. W. Thomas,
                         preprint ADP-92-192-T120.}
\REF\COMMA{See Refs. \MQ ~ and \CQR~ for obtaining the expression
           of $K$. }
\REF\SIZE{R. C. Barrett and D. F. Jackson,
               {\it Nuclear Sizes and Structure}
                \rm (Clarendon, Oxford, 1977).}
\REF\AP{G. Altarelli and G. Parisi,
                            Nucl. Phys. {\bf B126}, 298 (1977);
        In deriving Eq. (2), we should note a factor of 2
        coming from averages over initial spin states.}
\REF\YUD{F. J. Yndur\`ain, {\it Quantum Chromodynamics}, \rm
                         (Springer-Verlag, New York, 1983).}
\REF\RALSTON{J. P. Ralston, Phys. Lett. {\bf 172B}, 430 (1986). }
\REF\RGR{For a recent summary of experimental data for deep
                inelastic scatterings, see
            R. G. Roberts and M. R. Whally,
                         J. Phys. G {\bf 17}, D1 (1991).}
\REF\SLACA{R. G. Arnold et al.,
                            Phys. Rev. Lett. {\bf 52}, 727 (1984). }
\REF\EMCB{J. Ashman et al.  (EMC collaboration),
                            Phys. Lett. {\bf 202B}, 603 (1988). }
\REF\LLE{C. H. Llewellyn Smith, Nucl. Phys. {\bf A434}, 35c (1985).}
\REF\MRS{A. D. Martin, R. G. Roberts, and W. J. Stirling,
                    Phys. Rev. {\bf D37}, 1161 (1988).}
\REF\KMRS{J. Kwiecinski, A. D. Martin, W. J. Stirling,
           and R. G. Roberts, Phys. Rev. {\bf D42}, 3645 (1990).}
\REF\BM{R. P. Bickerstaff and G. A. Miller,
                      Phys. Lett. {\bf 168B}, 409 (1986).}
\REF\KUMA{S. Kumano and J. T. Londergan,
                 Comp. Phys. Commu. {\bf 69}, 373 (1992).}
\REF\QIUCOM{J. Qiu, personal communications (1992);
                Using $K[p_1(x_1)p_2(x_2)+p_1(x_2)p_2(x_1)]\Rightarrow
                       2Kp_1(x_1)p_2(x_2)$ in Eqs. (A3.3) and (A3.4),
                we obtain Eq. (24) of Ref. \CQR ~ from
                Eqs. (A3.1)$-$(A3.4) with $x_1\leftrightarrow x_2$
                in the last integral of Eq. (A3.2) .}
\REF\SKGLUE{S. Kumano, Phys. Lett. {\bf 298B}, 171 (1993);
            This paper is based on the model in section 3.}
\FIG\FIGONE{Schematic pictures of parton recombination processes.}
\FIG\FIGTWO{Parton fusions for (a) $qG \rightarrow q$,
                               (b) $q\bar q \rightarrow G$,
                               (c) $Gq \rightarrow q$, and
                               (d) $GG \rightarrow G$.}
\FIG\FIGTHREE{Momentum cutoff for leak-out partons,
              $w(x)=exp(-m_{_N}^2 z_0^2 x^2/2)$.}
\FIG\FIGFOUR{Comparisons with (a) SLAC data [\SLACA],
                              (b) EMC-90 [\EMCA]
                                 and NMC [\NMC] data for Ca.
             Solid (dotted) curves are obtained by using
             the MRS-1 (KMRS-B0) input distributions.
             $Q^2$=5 GeV$^2$ and $z_0$=2 fm.
             (A) recombinations,
             (B) recombinations + rescaling ($\xi_{_A}=1.86$).
             (C) recombinations with gluon-shadowing effects,
             (C$'$) the same as C except for the rescaling.
             The only KMRS-B0 curve is shown in C$'$.
             See text for detailed explanations of A, B, C, and C$'$.}
\FIG\FIGFIVE{(a) MRS gluon distributions.
                 See Ref. \MRS ~
                 and text for details of
                 hard, soft, and $1/\sqrt {x}$ gluon distributions.
            (b) Comparisons with E665 data for Xe [\ESIX]
                                 and EMC-88 data for Sn [\EMCB].
            Theoretical results are obtained (b) for the Xe nucleus
            ($\xi_{_A}=2.24$).
             $Q^2$=4 GeV$^2$ and $z_0$=2 fm. }
\FIG\FIGSIX{Comparisons with (a) SLAC data [\SLACA],
                               (b) NMC  data [\NMC] for He.
              $Q^2$=0.8, 5, and 20 GeV$^2$ for the dotted,
              solid, and dashed curves respectively.
              $z_0$=2 fm, $\xi_{_A}^{^V}=1.43$
               and $Q_0^{~2}$=0.8 GeV$^2$.}
\FIG\FIGSEVEN{Comparisons with (a) SLAC [\SLACA] and EMC-88 [\EMCB] data,
                               (b) EMC-90 [\EMCA]
                                   and NMC [\NMC] data for C.
             Notations for the dotted, dashed, and solid curves
             are the same in Fig. 6.
             $\xi_{_A}^{^V}=1.60$.}
\FIG\FIGEIGHT{Comparisons with (a) SLAC data [\SLACA],
                              (b) EMC-90 [\EMCA]
                                 and NMC [\NMC] data for Ca.
                               $\xi_{_A}^{^V}=1.86$.}
\FIG\FIGNINE{Comparisons with (a) SLAC data for Ag [\SLACA],
                                 EMC-88 data for Sn [\EMCB],
                                 and some of E665 data for Xe [\ESIX],
                             (b) E665 data for Xe [\ESIX]
                                 and EMC-88 data for Sn [\EMCB].
             Calculated results are for Ag
               ($\xi_{_A}^{^V}=2.17$) in (a)
             and for Xe ($\xi_{_A}^{^V}=2.24$) in (b).}
\FIG\FIGTEN{Dependence on $Q_0^{~2}$. $Q_0^{~2}$=2.0 GeV$^2$ is taken.
              $Q^2$=2, 5, and 20 GeV$^2$ for the dotted,
              solid, and dashed curves respectively.
              $z_0$=2 fm and $\xi_{_A}^{^V}=1.86$.}
\FIG\FIGELEVEN{Sea-quark and gluon distributions of
               MRS-1 (solid curves) and KMRS-B0 (dotted curves).}
\FIG\FIGTWELVE{Dependence on input distributions (see Fig. 11).
               Results by using the KMRS-B0 inputs
               for (a) $^{12}$C and (b) $^{40}$Ca.
              $Q^2$=0.8, 5, and 20 GeV$^2$ for the dotted,
              solid, and dashed curves respectively.
              $z_0$=2 fm. $\xi_{_A}^{^V}$=1.60 for C and 1.86 for Ca.
               $Q_0^{~2}$=0.8 GeV$^2$.
               Compare them with the results in Figs. 7b and 8b.}
\FIG\FIGTHIRTEEN{Dependence on the cutoff $z_0$.
                 Only $Q^2$=5 GeV$^2$ curves
                 are shown. $z_0$=2.0 (3.0) fm for the solid (dashed)
                 curves. $\xi_{_A}^{^V}=1.86$ and $Q_0^{~2}$=0.8 GeV$^2$.}
\FIG\FIGFOURTEEN{Rescaling are used for all partons.
                 Notations and parameters are same in Fig. 7b.}
%
\Pubnum{\twelverm MKPH-T-93-04, IU/NTC 92-20}
\date={March 24, 1993}
\titlepage

$~~$

$~~$

\title{{\bf Nuclear Shadowing in a Parton Recombination Model}}

\vskip 0.6cm

\author{S. Kumano $^*$}

\medskip

\address{Institut f\"ur Kernphysik, Universit\"at Mainz}
\vskip -0.2cm
\address{6500 Mainz, Germany}
\vskip -0.2cm
\address{and}
\vskip -0.2cm
\address{Nuclear Theory Center, Indiana University}
\vskip -0.2cm
\address{Bloomington, Indiana 47408, U.S.A.}

\medskip

$~$

\abstract
Deep inelastic structure functions $F_2^A(x)$
are investigated in a $Q^2$ rescaling model with
parton recombination effects.
We find that the model can explain
experimentally measured $F_2^A(x)$ structure functions
reasonably well
in the wide Bjorken$-x$ range ($0.005<x<0.8$).
In the very small $x$ region ($x<0.02$),
recombination results are very sensitive to input
sea-quark and gluon distributions.

$~$

$~$

\vfill
\noindent
\hrule width6cm

\noindent
* address after April 1, 1993:
  Department of Physics, Saga University, Saga 840, Japan.

$~~~$

\noindent
{submitted to Phys. Rev. C
   \hfill PACS numbers: 13.60.Hb, 25.30.-c}

\endpage
\sequentialequations
%

\centerline{\bf{1. Introduction}}

Several years have passed since the discovery
of nuclear modifications in structure functions
$F_2(x)$ by the European Muon Collaboration
(``old'' EMC effect) [\EMCX].
In spite of initial expectation of an explicit
quark signature in nuclear physics,
it is still not clear whether the effect originates in
nucleon substructures
or just in nuclear physics.
There are several proposed models for explaining the EMC effect.
Some models tried to interpret it by explicit quark
effects, e.g. $Q^2$ rescaling models [\RESCALE]
and six-quark bag models.
Others investigated models based on
conventional nuclear physics, such as nuclear binding
and pion excess.
For details of these models, we suggest that interested readers look
at summary papers in Ref. \ARGONNE.
Although there are still experimental activities
[\FNAL] for investigating these different models,
it is rather difficult
to discriminate among these models.

Most of these investigations discuss a
``global'' EMC effect in the sense that the effect is
averaged over all constituents in a nucleus.
However, it is interesting to investigate possible
semi-inclusive or semi-exclusive processes
for finding a ``local'' EMC effect [\SKFEC]
and possible relations between a local gluonic
EMC effect and the $J/\psi$ suppression [\SK].

Considering the fact that the average nucleon separation
in nuclei is 2.2 fm and the nucleon diameter is 1.8 fm,
we expect that nucleons in nuclei could overlap strongly.
If a multiquark cluster is formed in a nucleus due to this
overlap, the confinement radius for such a quark should be different
from the one in a free nucleon. Using this kind of simple picture
in a small $Q^2$ region and the $Q^2$-evolution equation
to compare with experimental data  at large $Q^2$,
we obtain a simple prescription
for the nuclear structure function $F_2^A (x,Q^2)$ [\RESCALE].
It is related to the nucleon
structure function by a simple $Q^2$ rescaling,
$F_2^A(x,Q^2) = F_2^N (x, \xi_{_A} Q^2)$, where
$\xi_{_A}$ is called as rescaling parameter.
Although the simple size change could be too simple to explain
many details of nuclear physics [\MONIZ], it is
a useful effective model in explaining deep inelastic data
in the medium $x$ region.

The New Muon Collaboration (NMC) [\NMC]
and the Fermilab E665 collaboration [\ESIX] recently measured
accurately the structure functions
$F_2(x)$ at very small $x$.
These data as well as EMC data [\EMCA]
provide an opportunity for investigating physics details
in the shadowing region ($x<0.1$).
The shadowing means that
central constituents are shielded due to surface constituents,
hence the cross section behaves like $A^\alpha$ ($\alpha<1$).
There are different ideas for explaining the shadowing
phenomena: vector meson dominance [\VECTOR],
parton recombination [\RECOMB $-$\CQR],
$q \bar q$ fluctuations of the virtual $\gamma$ [\FF],
pomeron dominance [\POMERON] (with Pauli blocking [\PRE]),
and hybrid models [\MIX].

At first, there are models based on the traditional
idea, the vector meson dominance model [\VECTOR].
A virtual photon transforms into
vector meson states
($\rho$, $\omega$, $\phi$),
which then interacts with a target nucleus.
The propagation length of the hadronic (H)
fluctuations is given by
$\lambda \approx 1 /| E_H - E_\gamma | =2 \nu /(M_H^2+Q^2)
         \approx 0.2/x$ fm.
For $x<0.1$, the propagation length ($>$2 fm)
exceeds the average nucleon separation (2.2 fm) in nuclei
and the shadowing takes place due
to multiple scatterings.
For example, the vector meson interacts
elastically with a surface nucleon and then
interacts inelastically with a central nucleon.
Because this amplitude is opposite in phase to
a one-step amplitude for
an inelastic interaction with the central nucleon,
the nucleon sees a reduced hadronic flux (namely the shadowing).

The parton recombination model [\RECOMB $-$\CQR] has been
investigated as a mechanism
for explaining the shadowing within the framework
of a quark-parton model.
In an infinite momentum frame,
the average longitudinal nucleon
separation in a Lorentz contracted nucleus is
$L = (2.2~ fm) M_A /P_A
     = (2.2~ fm) m_{_N}/p_{_N}$,
and the longitudinal localization size of a parton with momentum $xp_{_N}$
is $\Delta L=1/(xp_{_N})$.
If the parton dimension ($\Delta L$)
exceeds the average nucleon separation ($L$),
partons from different nucleons could interact
(we call this parton recombination or parton fusion) significantly
and the shadowing could occur due to processes
in Figs. 1b, 1c, 1e, and 1f.
Partons with momentum fraction $x$ are lost in these processes,
so that their contributions
are negative $\Delta F_2(x)<0$
(shadowing). However, $\Delta F_2(x)$ depends
much on input sea-quark and gluon distributions
as we find in sections 3 and 4.
A relevant $x$ region for the shadowing is
obtained by using $L < \Delta L$ and
we find $x<0.1$.
Even though the recombinations
could produce shadowing type
effects in the small $x$ region, they are very small
compared with experimental shadowings
if they are calculated at $Q^2$=5 GeV$^2$.
On the other hand, modifications of $F_2(x)$ due to recombination
effects on gluon distributions are large [\CQR].
Details of numerical results are discussed in section 3.

Because the recombination results by Close, Qiu, and Roberts
[\CQR] are not compared in detail with the recent experimental
data, we investigate whether the model
can explain the EMC, NMC, and E665 shadowing data
in this investigation. Furthermore,
we study whether the $Q^2$ rescaling model
combined with parton recombination effects [\CR,\PREDA]
can explain
the $F_2(x)$ structure functions in the whole Bjorken$-x$ range
[\LIUTI].

Earlier comparisons with experimental data
have been made in similar parton pictures [\PREDA].
Let us clarify differences from and advantages over
the previous investigations.
(1) Covolan, Predazzi, and others [\PREDA]
   used the Qiu's parametrization [\MQ] for the shadowing
   and combined its results with the rescaling.
   We note that the $Q^2$ evolution is well investigated
   in Ref. \MQ; however, the $x$ dependence is rather
   assumed. Therefore, we cannot have a dynamically consistent
   picture by using the shadowing-parametrization
   with the rescaling.
   The $x$ dependence of the shadowing was later investigated
   in Ref. \CQR. Using this shadowing picture combined
   with the rescaling, we could possibly
   obtain a dynamically-consistent parton model
   in the wide $x$-range.
   This is the purpose of our investigation.
(2) We study $Q^2$ evolution effects
    in the combined model.
(3) We investigate effects of input sea-quark and
    gluon distributions on the shadowing.
(4) Improved experimental data became available by the NMC and E665.

$~~~~$

\centerline
{\bf{2. Parton Recombinations}}

If a nucleon is in a nucleus, parton distributions
are modified because of the existence of neighboring nucleons.
Partons from different nucleons could interact with each other,
and the interactions become important, especially
in the shadowing region.
Processes contribute to modifications
in a quark distribution $q(x)$ are shown in
Fig. 1. For example, Fig. 1a indicates that
a quark from the nucleon 1 fuses with
a gluon from the nucleon 2 and produces a quark
with momentum $x$.
Because the process creates
a quark with the momentum $x$, this is a positive contribution
to the quark distribution at $x$.
There are five other contributions as shown in Fig. 1.
In general, a modification of a parton distribution
$p_3(x_3)$, due to the
process of producing the parton $p_3$ with the momentum $x_3$
by a fusion of partons $p_1$ and $p_2$, is given by [\CQR]
$$
\Delta p_3 (x_3) = K \int dx_1 dx_2
                 ~p_1 (x_1)~ p_2 (x_2) ~
\Gamma_{p_1 p_2 \rightarrow p_3} (x_1,x_2,x_3=x_1+x_2)
{}~\delta (x,x_1,x_2)~~,
\eqno(1)
$$
where $K$ is given by $K=9A^{1/3}\alpha_s/(2R_0^2 Q^2)$ [\COMMA].
Nuclear radius $R$ is $R=R_0 A^{1/3}$ with $R_0$=1.1 fm [\SIZE]
and the strong-interaction coupling constant is
$\alpha_s (Q^2) = 4 \pi/ [\beta_0 ln (Q^2/\Lambda^2)]$
with $\beta_0 =11-2N_f/3$, $N_f$=3.
The momentum conserving $\delta$ function is given by
$\delta (x-x_1-x_2)$ for Figs. 1a and 1d and
$\delta (x-x_1)$     for Figs. 1b, 1c, 1e, and 1f.
The parton fusion function $\Gamma(x_1,x_2,x_3)$
is a probability for producing a parton $p_3$
with momentum $x_3$ by a fusion of partons $p_1$ and $p_2$
with momenta $x_1$ and $x_2$ respectively.
Four possible parton fusion processes are shown in Fig. 2.
It is related to a splitting function
$P_{p_1 p_3} (z)$ in the Altarelli-Parisi
equation [\AP] by
$$
\Gamma_{p_1 p_2 \rightarrow p_3} (x_1,x_2,x_3) ~=~
  {{x_1 x_2} \over {x_3^2}} ~
  P_{p_1 \leftarrow p_3} ({{x_1} \over {x_3}}) ~
  C_{p_1 p_2 \rightarrow p_3} ~~~.
\eqno(2)
$$
\noindent
The splitting function $P_{p_1 p_3}(z)$ is the probability
of finding the parton $p_1$ with fraction $z$ of
the parent momentum in the parton $p_3$.
$C_{p_1 p_2 \rightarrow p_3}$ is the ratio of color factors
in processes $p_1 p_2 \rightarrow p_3$
and $p_3 \rightarrow p_1 p_2$.
For example,
$\displaystyle{
C_{qG \rightarrow q}
=\sum_{(l,a),k}^{\textstyle{-}} (t_{kl}^a)^* t_{kl}^a
/\sum_{(k),l,a}^{\textstyle{-}} (t_{lk}^a)^* t_{lk}^a
=}
$
$
[C_F/(N_C^2-1)]/C_F
$,
where averages are taken over initial color indices
($k$ in the denominator,
 $l$ and $a$ in the numerator).
$t_{lk}^a$ is given by the $SU(3)_c$ Gell-Mann
matrix by $t_{lk}^a=\lambda_{lk}^a/2$, and $C_F$ is
$C_F = (N_C^2-1)/(2N_C)$ [\YUD].
In this way, the color factors are calculated as:

\noindent
$
{}~~~~~~~~~~~~~~~
\displaystyle{
C_{qG \rightarrow q}={ 1 \over {N_C^2-1}} = { 1 \over 8 } ~~~~~,~~~~~
C_{q\bar q \rightarrow G} ={ {C_F} \over {N_C T_F}} = { 8 \over 9} ~~~,
}
$

\vskip 0.4cm
\noindent
$
{}~~~~~~~~~~~~~~~
\displaystyle{
C_{Gq \rightarrow q} = {{ T_F} \over {N_C C_F}} = { 1 \over 8} ~~~~~,~~~~~
C_{GG \rightarrow G} = { 1 \over {N_C^2 -1}} = {1 \over 8 } ~~~,
}
$
\hfill$(3)$
\vskip 0.4cm

\noindent
where $T_F=1/2$ and $N_C=3$.
Using the above results and splitting functions
in Ref. \AP, we obtain
the parton fusion functions:

\vskip 0.4cm
\noindent
$
{}~~~~~~~~~~~~~~~
\displaystyle{
\Gamma_{qG \rightarrow q'} (x_1,x_2,x_3) ~=~
                    {1 \over 6} ~z~ (1+z^2)   ~~~,
}
$

\noindent
$
{}~~~~~~~~~~~~~~~
\displaystyle{
\Gamma_{q \bar q \rightarrow G} (x_1, x_2, x_3) ~=~
                { 4 \over 9 } ~z~(1-z)~ [z^2 +(1-z)^2]  ~~~,
}
$

\noindent
$
{}~~~~~~~~~~~~~~~
\displaystyle{
\Gamma_{Gq \rightarrow q'} (x_1,x_2,x_3) ~=~
                {1 \over 6}~ (1-z)~ [1+(1-z)^2]    ~~~,
}
$
\hfill$(4)$

\noindent
$
{}~~~~~~~~~~~~~~~
\displaystyle{
\Gamma_{G_1 G_2 \rightarrow G_3} (x_1,x_2,x_3) ~=~
                {3 \over 4} ~z ~(1-z)~
                [ { z \over {1-z}} + {{1-z} \over z}
                  +z(1-z) ]                          ~~~,
}
$
\vskip 0.4cm

\noindent
where $x_3=x_1+x_2$ and $z \equiv x_1/x_3$.

Using these fusion functions, we calculate modifications
of quark distributions in nuclei for processes shown
in Fig. 1. Explicit expressions for the modifications $\Delta q_i (x)$
are shown in Appendix (Eq. (A1)). From Eq. (A1) and a similar
expression for $\Delta \bar q_i(x)$, we obtain recombination
contributions to the structure function $F_2(x)$
in Eqs. (A2.1)$-$(A2.4).
Direct recombination effects on the EMC effect are given
by the ratio
$$
R_A ({\rm recombination}) ~=~ 1~+~
{{ \Delta F_2^A (x, Q^2) } \over { F_2^D(x,Q^2) } }  ~~~,
\eqno(5)
$$
where $\Delta F_2 ^A (x,Q^2)$ is calculated
by using Eqs. (A2.1)$-$(A2.4) with
input parton distributions at $Q^2$.
The most important physics for the shadowing
in the recombination model
is the effect due to modifications in gluon distributions.
Recombination effects on the gluon distribution
are given in Eqs. (A3.1)$-$(A3.4).
Calculating these equations,
we find that the recombination mechanism produces
large shadowing effects
on the gluon distribution [\CQR].

$~~~$

\centerline
{\bf {3. Gluon-shadowing effects on $\bf F_2^A(x)$}}

\noindent
{\bf{3.1 An approximate way to take into account
         the gluon shadowing}}

Gluons and quarks are coupled to each other and their distributions
are related by the Altarelli-Parisi equation. Because of
the large modifications in $G(x<0.1)$ due to the recombinations,
we expect that
sea-quark distributions are also affected by the
gluon modifications. In the very small $x$ region,
the Altarelli-Parisi equation
is dominated by gluon dynamics, and it
can be solved analytically.
As a result, there is a relationship between
a sea quark and gluon distributions [\RALSTON]:
$
\displaystyle{
x q_i^{sea} (x) = - { x \over {12}}
                    { \partial \over {\partial x}}
                    [x G(x)]
}$, where $i$=u, d, or s.
Using this relation, we obtain a contribution
to $F_2(x)$
from the modification in
the gluon distribution:
$$
\delta F_2 (x,Q^2) = - \theta (x_0-x) ~ {4 \over 3}~ {x \over {12}} ~
                 {\partial \over {\partial x}}
                 [ x \Delta G(x,Q^2) ] ~~~,
\eqno(6)
$$
where $x\Delta G(x,Q^2)$ is given in Eqs. (A3.1)$-$(A3.4), and
a step function $\theta (x_0-x)$ is introduced
because the relation is valid only at very small $x$.
It is defined by $\theta (x_0-x) = 1$ (for $x<x_0$)
                                  or 0   (for $x>x_0$).
We should note that this is a crude estimate
of the gluonic contribution.
To be precise, the recombination calculations
are done at small $Q^2$, then
the results are evolved to the $Q^2$ region
where experiments were done, by using a
nuclear-modified Altarelli-Parisi equation [\MQ].
This approach is discussed in section 4.
Combining all the contributions, we obtain
nuclear structure functions $F_2^A(x)$ calculated
in our parton model (the $Q^2$ rescaling model
with parton recombination effects):

\vskip 0.4cm
\noindent
$
{}~~~~~~~~~~~~~~~
\displaystyle{
{{F_2^A(x,Q^2)} \over {F_2^D(x,Q^2)}} ~=~
{{\tilde F_2^A (x,Q^2) + \Delta F_2^A (x, Q^2)
                       + \delta F_2^A (x, Q^2)} \over
 {F_2^D(x,Q^2)} }        ~~~,
}$
\hfill$(7)$
\vskip 0.4cm

\noindent
where
$\tilde F_2^A$ is given  by $\tilde F_2^A(x,Q^2)$=
$F_2^N(x,\xi_{_A} Q^2)$ in the rescaling model.

$~~~$

\noindent
{\bf{3.2 Results by using the analytical solution at small $x$}}

We calculate nuclear structure functions
and investigate whether our theoretical results are
compatible with experimental data [\RGR]
by a SLAC group [\SLACA],
EMC [\EMCA,\EMCB], NMC [\NMC], and E665 [\ESIX].
It is interesting to investigate not only
the shadowing region at small $x$ but also
the large $x$ region,
which is traditionally called ``nucleon-Fermi-motion'' part.

In evaluating equations in Appendix, we first assume that
a leak-out parton is a sea quark or a gluon and that
the momentum cutoff function [\CQR,\LLE]
for this parton is taken as
$w(x)=exp(-m_{_N}^2 z_0^2 x^2/2)$, namely
$q^*(x)=w(x) q_{sea} (x)$, $\bar q^{~*}(x)=w(x) \bar q (x)$,
and $G^* (x)=w(x) G(x)$.
Input parton distributions are given by a recent parametrization
by Martin, Roberts, and Stirling (MRS) [\MRS]
or by Kwiecinski, Martin, Roberts, and Stirling (KMRS) [\KMRS].
$Q^2=5~\rm (or~ 4) ~GeV^2$ is used
in the input parton distributions
and for calculating $K$ in Eq. (1).
The cutoff function $w(x)$ supposedly takes into account
effects of bound partons and it is shown as the function
of $x$ in Fig. 3.
As discussed in the introduction, $x=0.1$ corresponds
to the length scale 2 fm, the
average nucleon separation in nuclei.
Hence, we expect that only partons $p^*$ with
$x {< \atop \sim} 0.1$ could
participate in the recombinations.
We then find in Fig. 3 that an appropriate choice is
$z_0$=2$-$3 fm.
We show theoretical results with $z_0$=2 fm in this section;
some $z_0$ dependencies are discussed in section 4.3.
$x_0$ in Eq. (6) is taken as $x_0=0.1$.
We do not show the ratio in Eq. (7) in the $x$
range $0.1<x<0.2$ because physics is not well
described by the crude estimate in Eq. (6)
for taking into account the gluonic modifications.
This $x$ region is investigated by a
more-complete $Q^2$-evolution picture
in section 4.
The QCD scale parameter $\Lambda$ in $\alpha_s (Q^2)$
is taken as $\Lambda$=0.2 GeV.

In our theoretical analysis, Ag, Sn, and Xe targets are
assumed as $^{107}$Ag, $^{118}$Sn, and
           $^{131}$Xe nuclei.
We take the rescaling parameters in Table II
of the Close, Jaffe, Roberts, and Ross paper [\RESCALE].
They are $\xi_{_A}$=1.43 ($^4$He),
                    1.60 ($^{12}$C),
                    1.86 ($^{40}$Ca),
                    2.17 ($^{107}$Ag), and
                    2.24 ($^{118}$Sn).
$\xi_{_A}$=2.24 is assumed for the $^{131}$Xe nucleus.
The above parameters are obtained at $Q^2$=20 GeV$^2$.
Even though the model can explain the data,
it is known that
the original fits are no longer valid if we include
Fermi-motion effects [\BM].
Furthermore, the rescaling parameters
are calculated in a ``semi-classical'' way in the
sense that overlapping volumes are estimated simply
by the geometrical ones.
Considering these points,
we do not think that the parameters are not well
defined in their overall magnitudes.
According to the dynamical rescaling model by
Close, Roberts, and Ross [\RESCALE]
in 1988,
nuclear dependence is the result of
a scale change. So, we could in principle set
a rescaling parameter by fitting $F_2^A(x)$
data of a medium-size nucleus, and then use
the density dependence of the rescaling model
in calculating $\xi_{_A}$ for other nuclei.
However, we find in our numerical analysis
that the above
$\xi_{_A}$ explain the SLAC data reasonably well,
so that we simply keep the original rescaling
parameters in Table II of Ref. \RESCALE.

We discuss results for the calcium nucleus
in Figs. 4a and 4b. In Fig. 4a,
theoretical results in Eqs. (5) and (7)
are compared with the SLAC data [\SLACA]
in the linear $x$ scale.
For simplicity, the deuteron structure functions
are assumed as
$F_2^D(x)=[F_2^p(x)+F_2^n (x)]/2\equiv F_2^N(x)$
in our theoretical analysis.
Two sets of results are shown: the solid curves
are obtained by using the MRS-1 distributions [\MRS]
and the dotted curves are by the KMRS-B0 [\KMRS].
A, B, C, and C$'$ in the figure show

$
{}~~~~~(A)~~~~~ [F_2^N(x,Q^2)+\Delta F_2(x,Q^2)]/F_2^N(x,Q^2)~~~,
$

$
{}~~~~~(B)~~~~~ [F_2^N(x,\xi_{_A} Q^2)+\Delta F_2(x,Q^2)]/F_2^N(x,Q^2)~~~,
$

$
{}~~~~~(C)~~~~~ [F_2^N(x,Q^2)+\Delta F_2(x,Q^2)+\delta F_2(x,Q^2)]
                    /F_2^N(x,Q^2)~~~,
$

$
{}~~~~~(C')~~~~~ [F_2^N(x,\xi_{_A} Q^2)
                   +\Delta \tilde F_2(x,Q^2)
                   +\delta \tilde F_2(x,Q^2)]
                    /F_2^N(x,Q^2)~~~,
$

\noindent
where $\Delta \tilde F_2(x,Q^2)$
  and $\delta \tilde F_2(x,Q^2)$ are recombination
results with $Q^2$-rescaled input distributions.
Direct recombination contributions,
shown by the curves with A, are positive
in the medium and large $x$ regions.
These come from the quark-gluon
fusion processes in Figs. 1a and 1d.
Combined results with the $Q^2$ rescaling
are shown by the curves with B.
The C and C$'$ curves are explained
in the following paragraph.
Fig. 4a indicates that our model
results are in overall agreement with
the experimental data.
The EMC effect in the medium $x$ region
is mainly due to the $Q^2$ rescaling in our model,
because the recombination contributions are
rather small.
It is noteworthy in these figures
that the recombination can explain
nuclear structure functions in the large $x$ region
without explicit Fermi-motion effects.
Physics in this region is usually attributed to
the nucleon Fermi motion in the nucleus.
In our model, a quark with $x {<\atop\sim} 1$ could
be pushed by a gluon from a different nucleon
and becomes a quark with $x>1$. This is the reason why
the ratio goes to infinity at $x \rightarrow 1$ in Fig. 4a.
Perhaps, there exists a correlation between these two
apparently different descriptions: the parton recombination and
the nucleon Fermi motion.

In Fig. 4b, theoretical results
are compared with EMC [\EMCA] and NMC [\NMC] data
in the logarithmic $x$ scale.
In the small $x$ region, the processes in Figs. 1b, 1c, 1e, and 1f
significantly
contribute and produce negative (shadowing) results
in the case of KMRS-B0.
However, they are too small to explain the whole shadowing data
if the recombinations are calculated at $Q^2$=5 GeV$^2$.
The important point in our model is the shadowing
which is produced through the gluon distributions.
Combined results are shown by
the curves with C
or C$'$ in Fig. 4b.
The figure indicates that the shadowing
produced through modifications
in gluon distributions can explain the EMC and NMC shadowing
data fairly well, even though the direct recombination effects
on $F_2(x)$ (shown by A) are very small.
The curves C are obtained without rescaling and
C$'$ is with the $Q^2$-rescaled inputs, where the same $\xi_{_A}$ is
used in all parton distributions.
We find that the differences between the C and C$'$ curves
are rather small considering the fact
that the $Q^2$ rescaling produces large sea-quark modifications.
This is because the $Q^2$-rescaled sea-quark distributions
and contributions (Eq. (6)) from $Q^2$-rescaled gluon
distributions almost cancel each other.

We have shown that our model can explain the NMC shadowing
in Fig. 4b.
However, the results are very sensitive to
input gluon distributions in the small $x$ region.
Because the E665 collaboration measured
$F_2^{Xe}(x)/F_2^{D}(x)$ at very small $x$ recently,
we compare our results
with the E665 data.
In order to find details of
input gluon distribution effects, we use parton distributions
with ``hard'', ``soft'', and ``$1/\sqrt{x}$'' gluon distributions
given by the MRS [\MRS]. Analytical
distributions at $Q^2$=4 GeV$^2$ are used as the input
distributions. The MRS soft, hard, and $1/\sqrt{x}$
gluon distributions are given by
$\delta_{_G}=0,~\eta_{_G}=5,~\gamma_{_G}=0$,
$\delta_{_G}=0,~\eta_{_G}=4,~\gamma_{_G}=9$, and
$\delta_{_G}=-1/2,~\eta_{_G}=4,~\gamma_{_G}=9$
respectively for the notation
$xG(x)=A_G x^\delta_{_G} (1-x)^{\eta_{_G}} (1+\gamma_{_G} x)$.
These gluon distributions are shown in Fig. 5a
and they are very different especially at $x<0.01$.
Our shadowing results are shown in Fig. 5b.
Because the experimental $Q^2$ of the E665 data
becomes very small ($Q^2<<1$ GeV$^2$) at $x<0.001$,
our perturbative results cannot be compared with the E665 data.
This is the reason why curves in the region $x<0.001$
are shown by the dashed ones.
As shown in the figure, our
calculations produce widely-different shadowing
results depending on the input gluon distribution.
If it is the $1/\sqrt{x}$ distribution,
shadowings are too large and are contradictory
to the experimental data. If it is the hard one,
theoretical results underestimate the shadowing.
Our theoretical results are nearly consistent with
experimental data if the input distributions
are the soft MRS distributions
at $Q^2$=4 GeV$^2$.

We found that our model with the recombination
and the rescaling works in the wide-$x$ region,
if an appropriate input gluon distribution is taken.
Nevertheless, it is not a complete description in the
sense that $F_2^A(x)$ in the $x\sim 0.1$ region
cannot be calculated. Furthermore, a curve
in the $x>0.2$ region and one in the $x<0.1$ are
not smoothly connected in Fig. 4b.
It implies that the analytical solution Eq. (6) at $x<0.1$
is used beyond the applicable region.
In order to have detailed comparisons
with experimental data and
to have a dynamically consistent picture,
$Q^2$ evolution effects should be fully
investigated. We address this point in
the following section.

$~~~$

\vfill\eject
\centerline
{\bf {4. $\bf Q^2$ evolution of nuclear structure functions}}

\noindent
{\bf{4.1 $\bf Q^2$ evolution}}

The approximate description with Eq. (6) is good enough
for rough magnitude estimates.
For detailed comparisons with the experimental data in the
$x\approx 0.1$ region,
we should solve $Q^2$-evolution equations.
Namely, we calculate the recombination and the rescaling
at small $Q^2$ ($\equiv Q_0^{~2}$), then obtained distributions
are evolved to those at larger $Q^2$ where experimental
data are taken. However, this is not an easy work
by the following reasons.

First, the shadowing data are mostly taken in the
$Q^2$ region of a few GeV$^2$.
We need to assign a very small value to $Q_0^{~2}$, for example
$Q_0^{~2}$=1 GeV$^2$, in order to
study evolution effects by comparing
our results with the existing data.
The perturbative QCD, especially in the leading order,
would become questionable in such small $Q^2$ region.
Nevertheless, we present the evolution effects
in this report because the gluon shadowing connected
with the $Q^2$ evolution is the essential part
of our model in explaining the $F_2$ shadowing,
as we found in the previous section.
Next-to-leading (and higher) order corrections are important
in the small $Q^2$ and small $x$ region.
Hence, our investigation should be considered as a first step
for understanding $F_2^A (x)$ dynamically
in the wide $x$ region
within the framework of a quark-parton model.

Secondly, the evolution equation for parton distributions in
a nucleus is given in Ref.~ \MQ.
Finding a numerical solution for the equation
is by itself a significant research problem.
So, instead of solving the nuclear evolution
equation, we use the Altarelli-Parisi equation
for our $Q^2$ evolution.
Both evolution results are qualitatively similar; however,
the nuclear-evolution shows smaller $Q^2$-dependence
for large nuclei [\MQ]. Because our results indicate
small $Q^2$ dependence for large nuclei in section 4.2,
this problem is not considered to be very serious.

We calculate $F_2^A(x)$ in the following way.
Valence-quark distributions in the nucleon are modified
according to the $Q^2$ rescaling mechanism at
$Q_0^{~2}$:
$\tilde V(x,Q_0^{~2})=V(x,\xi_A^V Q_0^{~2})$.
In order to satisfy the momentum conservation
$\displaystyle{\int dx x[\tilde u_v(x)+\tilde d_v(x)
                       +\tilde S(x)+\tilde G(x)]=1}$,
sea-quark and gluon distributions
are simply modified as $\tilde S(x,Q_0^{~2})=C_{sg}S(x,Q_0^{~2})$
           and $\tilde G(x,Q_0^{~2})=C_{sg}G(x,Q_0^{~2})$.
(It is not clear whether the rescaling picture
could be used also for sea-quarks and gluons.
This issue is discussed in section 4.3.)
Obtained parton distributions are then used for
calculating the recombination effects in Appendix, and
we get nuclear parton distributions at $Q_0^{~2}$.
These distributions are evolved to those at larger
$Q^2$ by using the ordinary Altarelli-Parisi equation [\AP].
For the flavor-singlet case, it is

$
\displaystyle{
{}~~~~~~~~
{\partial \over {\partial t}}  q_s(x,t) =
     \int_x^1 {{dy} \over y}
     ~[~ q_s(y,t) ~P_{qq}({x \over y})
       ~+~ G (y,t) ~P_{qG}({x \over y}) ~]
{}~~~,
}
$

$
\displaystyle{
{}~~~~~~~~
{\partial \over {\partial t}}  G (x,t) =
     \int_x^1 {{dy} \over y}
     ~[~ q_s(y,t) ~P_{Gq}({x \over y})
       ~+~ G (y,t) ~P_{GG}({x \over y}) ~]
{}~~~,
}
$
\hfill (8)

\noindent
where $q_s$ is the flavor-singlet distribution
given by $\displaystyle{q_s=\sum_i (q_i+\bar q_i)}$,
$P_{ij}$ is the splitting function,
and $t$ is defined by
$t=-(2/\beta_0) ln[\alpha_s(Q^2)/\alpha_s(Q_0^{~2})]$.
We used a solution of this integrodifferential equation
in the leading order [\KUMA].
In the flavor-nonsinglet case, the gluon does not enter
into the evolution equation:
$$
{\partial \over {\partial t}}  q_{ns} (x,t) =
     \int_x^1 {{dy} \over y}
     ~q_{ns}(y,t) P_{qq}({x \over y})
{}~~~,
\eqno{(9)}
$$
where $q_{ns}=q_i-\bar q_i$.
We use the distributions at $Q_0^{~2}$
calculated by the rescaling
and the recombinations in the above equations:
$q_{ns}(x,Q_0^{~2})
  =\tilde u_v(x,Q_0^{~2})+\tilde d_v(x,Q_0^{~2})$,
$q_{s} (x,Q_0^{~2})
  =\tilde u_v(x,Q_0^{~2})+\tilde d_v(x,Q_0^{~2})+\tilde S(x,Q_0^{~2})$,
and $G(x,Q_0^{~2})=\tilde G(x,Q_0^{~2})$.
We then obtain evolved distributions by solving
the above equations,
and $F_2$ at $Q^2$ is given by
$F_2(x,Q^2)=[xq_{ns}(x,Q^2)+4xq_s(x,Q^2)]/18$.
The use of the above equations instead of
the nuclear evolution equations causes a problem
in the large $x$ region, because
the recombinations produce distributions at $x>1$.
However, it is not a serious problem in comparison
with the experimental data discussed in the next section.

$~~~$

\noindent
{\bf{4.2 $Q^2$ evolution results}}

Structure functions $F_2^A(x)$ are calculated in
the following way.
Because the parton distributions in the MRS
have simpler analytical forms than those
in the KMRS,
we first employ the MRS-1 (soft gluon)
as our parton distributions in the nucleon.
Using the analytical form of the MRS-1
at $Q^2$=4 GeV$^2$ and the evolution subroutine
in Ref. \KUMA,
we calculate distributions at $Q_0^{~2}(\approx 1 ~{\rm GeV}^2)$.
Then, the $Q^2$ rescaling is used for valence distributions
and the constant $C_{sg}$ is determined so that the
momentum conservation is satisfied.
Input distributions at $Q_0^{~2}$ are parametrized by
a simple analytical form,
$xp(x)=A x^\delta (1-x)^\eta (1+\gamma x)$.
Constants $A$, $\delta$, $\eta$, and $\gamma$
are obtained by fitting numerical results.
The recombinations are calculated by using
the modified input distributions at $Q_0^{~2}$,
and obtained results are again fitted by
$xp(x)=A x^\delta (1-x)^\eta (1+\gamma x)$.
For the valence quarks, we used
$xV(x)=A x^\delta (1-\beta x)^\eta (1+\gamma x)$,
where $\beta<1$, so that we could take into account
the fusion effects at large $x$ (=0.8$-$0.9).
The recombinations produce distributions
at $x>1$; however, we simply neglected
such distributions in our $Q^2$ evolution.
This causes slight inconsistency, but
the neglected effect on the momentum conservation
is of the order of 10$^{-4}$\%, which is insignificant.

We discuss results for the nuclei He, C, Ca,
Ag, Sn, and Xe in Figs. 6-9.
In Figs. 6a-9a, our theoretical results
are compared with SLAC [\SLACA]
and EMC [\EMCB] (E665 [\ESIX])
data in the linear $x$ scale.
In order to see the shadowing due to the $Q^2$ evolution,
we should choose $Q^2\sim$1 GeV$^2$
because the data were
taken in the region of a few GeV$^2$.
The value of $Q_0^{~2}$ is adjusted so that
our results agree with the shadowing data
for $^{40}$Ca, and we obtain $Q_0^{^2}$=0.8 GeV$^2$.
We fixed the cutoff at $z_0$=2 fm as it was used
in section 3.
We discuss dependence of our results
on these parameters in the next section.
Another problem is how we should
choose the rescaling parameters at very small $Q^2$.
As we did in section 3, we may simply determine
$\xi_{_A}$ for a medium-size nucleus and then use
the A-dependence for other nuclei.
However, it turns out that an original rescaling
parameter for a medium-size nucleus is not a bad choice,
so we simply use the parameters in Ref. \RESCALE.
Three curves with $Q^2$=0.8, 5.0, 20.0 GeV$^2$ are
shown in Figs. 6-9.
Figures 6a-9a indicate that our model can explain
the experimental data well except for
the $x\approx 0.7$ region of large nuclei.
The results in this region depend much on
the input valence distribution and
the cutoff (see section 4.3).
If we change the input to the KMRS-B0 distributions,
our calculations work very well for large nuclei,
but the EMC effect for $^4$He is overestimated.

In Figs. 6b-9b, theoretical results
are compared with EMC [\EMCA], NMC [\NMC],
and E665 [\ESIX] data in the logarithmic $x$ scale.
Our results are not shown at $x<0.001$ in Fig. 9b because
the used evolution subroutine [\KUMA] does not
have good accuracy in such region and our perturbative
calculations cannot be compared with the data
with very small $Q^2$.
We find in Figs. 6b-9b that our results agree quite well
with the experimental data.
In the small nuclei (Figs. 6b and 7b),
there are some $Q^2$ dependence at $Q^2=0.8-2$ GeV$^2$,
but it becomes $Q^2$ independent at $Q^2>$ a few GeV$^2$.
We have very small $Q^2$ dependence in the larger nuclei
as shown in Figs. 8b and 9b.
We note that rather
large $Q^2$ dependence in
Figs. 6b and 7b should not be taken seriously at this stage,
because small variations in input sea-quark and gluon
distributions change the dependence
drastically as we find in section 4.3.

$~~~$

\noindent
{\bf{4.3 Dependence on the parameters}}

We discuss how our results depend
on the parameters,
the input distributions, and the rescaling assumption.
We first check the choice of $Q_0^{~2}$.
We choose $Q_0^{~2}$=2.0 GeV$^2$
instead of 0.8 GeV$^2$ in Fig. 10.
Distributions at 2 GeV$^2$ are
calculated and are then evolved to those
at 5 and 20 GeV$^2$.
It is obvious that our calculations underestimate the experimental
shadowings and that the $Q^2$ dependence is very different
from the one in Fig. 8b.
The differences are due to the $Q^2$ factor
in $K$ (see Eq. (1)) and due to input sea-quark
and gluon distributions. In fact, we find in the
following that slight modifications in those distributions
change our shadowing results at $Q_0^{~2}$
significantly.

In order to check input-distribution effects, we take
the KMRS-B0 distributions which are
slightly different from those of the MRS-1.
These distributions are shown in Fig. 11.
Even though two distributions do not look very
different, $Q^2$ dependency of the KMRS-B0 results
for the carbon nucleus
in Fig. 12a is quite different from that in Fig. 7b.
The KMRS-B0 results for the calcium
in Fig. 12b are rather similar to
those of the MRS-1 in Fig. 8b.
Because the parton distributions in the small-$x$ region
are not well known, we inevitably have such uncertainty
in our model.

Effects of a momentum cutoff for leak-out partons are shown
in Figs. 13a and 13b.
{}From Fig. 3, we decided to take $z_0$=2 fm and we have been using
this cutoff so far.
Results with $z_0$=3 fm are shown by the dashed curves.
Our model with larger $z_0$ produces smaller recombination effects
at large $x$ and larger effects at small $x$. The agreement
with the data becomes better in Fig. 13a in the medium-$x$
region; however, it becomes worse in the small-$x$
region in Fig. 13b.
Because the cutoff is one of the important factors in our model,
efforts should be made to estimate it
in an independent way.

The last issue is the rescaling.
The original rescaling model is
intended for valence quarks.
It is not clear whether
sea-quarks and gluon follow the same rule
of the scale change in nuclei.
In our analysis of this section, we used the rescaling only for
valence quarks. Sea-quark and gluon
distributions are simply increased by a constant amount ($C_{sg}$)
so that they carry
a momentum deficit produced by the valence rescaling.
This is an arbitrary assumption.
We may choose other scenario, for example
the $Q^2$ rescaling for all partons.
Using this rescaling picture, we obtain the results
in Fig. 14. The shadowings are much underestimated
in this case.
Although numerical values do not agree completely
with those in Fig. 4b, we find a similar tendency
in the sense that
calculated shadowings become smaller
in the $x$ region ($0.01<x<0.1$)
if the rescaling is used for all partons.

We learned the following from the above analyses.
There are a few parameters and assumptions in
our model and obtained results depend
inevitably on their choice.
The major factors are
(1) $Q_0^{~2}$,
(2) input S(x) and G(x) in the small $x$,
(3) $z_0$,
and
(4) $Q^2$ rescaling for (valence) partons.
We find that magnitude of our shadowing and
the $Q^2$ dependence are very sensitive to the above
factors.
Nevertheless, we find that it is possible to choose
a set of the factors within our initial-expectation ranges
(e.g. $Q^2\sim$1 GeV$^2$, $z_0\sim$ 2 fm)
and we explain many existing data.
Obviously, much efforts should be done
for investigating these factors and
also next-to-leading order effects.

$~~~$

\centerline
{\bf{4. Conclusions}}

We investigated
nuclear structure functions $F_2(x,Q^2)$
from small $x(\approx 10^{-3})$ to large $x(\approx 0.9)$
in a parton model based
on a $Q^2$ rescaling model with parton recombination
effects in order to compare them with the recent experimental data.
As a result, we obtained reasonably
good agreement with the experimental
data in the region ($0.005<x<0.8$).
In the large $x$ region, the ratio ($F_2^A(x)/F_2^D(x)>1$)
is explained by quark-gluon recombinations, which produce
results similar to those by the nucleon Fermi motion.
In the medium $x$ region, the EMC effect is mainly due to
the $Q^2$ rescaling mechanism in our model. In the small $x$ region,
shadowing effects are obtained through modifications
in gluon distributions.
However, our shadowings at very small $x(<0.02)$ are very sensitive
to the input gluon distribution.
We have a few parameters in our model; however,
we could choose a set of the parameters and explain
many existing experimental data.
$Q^2$ variations of our shadowing results depend much
on the input sea-quark and gluon distributions, and also
on the parameters.


$~~~~~$

\centerline
{\bf{Acknowledgments}}

This research was supported by
the Deutsche Forschungsgemeinschaft (SFB 201)
and by the US-NSF under Contract No. NSF-PHY91-08036.
S.K. thanks Drs. F. E. Close, J. Qiu, and R. G. Roberts
for helpful suggestions
and for patiently answering questions
about parton recombinations;
Drs. M. van der Heijden, Q. Ingram, and
G. van Middelkoop for discussion about NMC experiments;
Drs. C. W. Salgado and H. M. Schellman for
information about the E665 experimental results;
Dr. W. J. Stirling for sending
their computer programs for calculating
the parton distributions in Ref. \KMRS.

\vfill\eject

\centerline
{\bf{Appendix}}

Although detailed formalisms are given in
Ref. \CQR, explicit equations used in this investigation
are shown in the following.
Using Eqs. (1)$-$(4) and changing the integration
variable $x_1 \leftrightarrow x_2$
for the process in Fig. 1e,
we obtain modifications of quark distributions
in nuclei due to the parton recombination mechanism:

\vskip 0.4cm
\noindent
$
\displaystyle{
x \cdot \Delta q_i (x) ~=~
  +   { K \over 6 } \int_0^x {{dx_2} \over {x_2}}
         ~x_2 G^*(x_2)~
        \biggl[~ (x-x_2) q_i(x-x_2)
     ~   \bigl\{1+( {{x-x_2} \over x} )^2 \bigr\}
}
$

\noindent
$
{}~~~~~~~~~~~~~~~
{}~~~~~~~~~~~~~~~
{}~~~~~~~~~~~~~~~
{}~~~~~~~~~~~
\displaystyle{  - ~xq_i(x)~ {x \over {x+x_2}}~
                       \bigl\{ 1+({x \over {x+x_2}} )^2 \bigr\}~ \biggr]
}
$

\noindent
$
{}~~~~~~~~~~~~~~~~~
\displaystyle{
- { K \over 6 } \int_x^1 {{dx_2} \over {x_2}} ~x q_i (x)~
         x_2 G^*(x_2) ~{ x \over {x+x_2}}~
         \bigl\{ 1+( {x \over {x+x_2}} )^2 \bigr\}
}
$

\noindent
$
{}~~~~~~~~~~~~~~~~~
\displaystyle{
- { {4K} \over 9 } x\int_0^1 dx_2  ~xq_i(x)~
         x_2 \bar q_i^*(x_2)~ {{x^2+x_2^2} \over {(x+x_2)^4}}
}
$

\noindent
$
{}~~~~~~~~~~~~~~~~~
\displaystyle{
+  { K \over 6 } \int_0^x {{dx_1} \over {x_1}}
       ~  x_1 G (x_1)~ \biggl [~ (x-x_1) q_i^*(x-x_1) ~
        \bigl \{ 1+( {{x-x_1} \over x} )^2 \bigr \}
}
$

\noindent
$
{}~~~~~~~~~~~~~~~
{}~~~~~~~~~~~~~~~
{}~~~~~~~~~~~~~~~
{}~~~~~~~~~~~
\displaystyle{
              -~xq_i^* (x)~ {x \over {x+x_1}}~
                   \bigl\{  1+({x \over {x+x_1}})^2 \bigr\} ~ \biggr ]
}
$

\noindent
$
{}~~~~~~~~~~~~~~~~~
\displaystyle{
- { K \over 6 } \int_x^1 {{dx_1} \over {x_1}}
        ~ x_1 G(x_1) ~x q_i^*(x)~ { x \over {x+x_1}}~
         \bigl \{ 1+( {x \over {x+x_1}} )^2 \bigr \}
}
$

\noindent
$
{}~~~~~~~~~~~~~~~~~
\displaystyle{
- { {4K} \over 9 } x \int_0^1 dx_2   ~ x q_i^* (x)
        ~ x_2 \bar q_i(x_2)~  {{x^2+x_2^2} \over {(x+x_2)^4}}
{}~~~~~,
}
$

\hfill$(A1)$

\noindent
where $*$ indicates a leak-out parton. For example
a gluon, which leaks out from the nucleon 2, interacts with
a quark in the nucleon 1 and produces a quark with momentum $x$
in the first term of Eq. (A1).
Explicit $Q^2$ dependence in parton distributions
is not shown in this appendix in order to simplify the notations.
Modifications in antiquark distributions are
obtained in the similar way by changing $q_i \leftrightarrow \bar q_i$
in Eq. (A1).
Using the above expression, we obtain
parton recombination effects on the structure functions $F_2(x)$ by
$\Delta F_2 (x) = {\displaystyle\sum_i}
                 e_i^2 x [ \Delta q_i (x)+ \Delta \bar q_i(x)]$

\vskip 0.4cm
\noindent
$
\displaystyle{
\Delta F_2 (x)
=\Delta F_2^{(1)} (x)
+\Delta F_2^{(2)} (x)
+\Delta F_2^{(3)} (x)
}
$
\hfill$(A2.1)$

$~~~$

\noindent
$
\displaystyle{
\Delta F_2^{(1)} (x) =
     +{ K \over 6 } \int_0^x {{dx_2} \over {x_2}}
        ~ x_2 G^*(x_2) ~
      \biggl[~ F_2 (x-x_2) ~
            \bigl \{ 1+( {{x-x_2} \over x} )^2 \bigr \}
}$

\noindent
$
\displaystyle{~~~~~~~~~~~~~~~~~~~~~~~~~~~~~
              ~~~~~~~~~~~~~~~~~
- ~F_2(x)~ {x \over {x+x_2}}~
    \bigl \{ 1+({x \over {x+x_2}})^2 \bigr \} ~ \biggr]
}
$

\noindent
$
{}~~~~~~~~~~~~~~~~~
\displaystyle{
- { K \over 6 } \int_x^1 {{dx_2} \over {x_2}}~ F_2 (x)
       ~  x_2 G^*(x_2) ~{ x \over {x+x_2}}~
         \bigl \{ 1+( {x \over {x+x_2} })^2 \bigr \}
}
$
\hfill$(A2.2)$

$~~~$

\noindent
$
\displaystyle{
\Delta F_2 ^{(2)} (x) =
       +  { K \over 6 } \int_0^x {{dx_1} \over {x_1}}
         ~x_1 G(x_1)~ \biggl [~
              \sum_i^{\textstyle{\textstyle{-}}}  e_i^2 ~
       \bigl \{ (x-x_1) q_i^*(x-x_1) + (x-x_1) \bar q_i^*(x-x_1) \bigr \}
}
$

\noindent
$
{}~~~~~~~~~~~~~~~~~~~~~~~~~~~~~~~~~~~~~~
{}~~~~~~~~~~~~~~~~~~~~~~~~~~~~~~~~~~~~~~
\displaystyle{
              \times  \bigl \{ 1+( {{x-x_1} \over x} )^2  \bigr \}
}
$

\noindent
$
{}~~~~~~~~~~~~~~~
{}~~~~~~~~~~~~~~~
{}~~~~~~~~~~~~~~~
\displaystyle{
                      - \sum_i^{\textstyle{-}}
                      e_i^2 ~
                \bigl \{ x q_i^*(x) + x \bar q_i^*(x) \bigr \}
                     ~  {x \over {x+x_1}}  ~
                \bigl \{ 1+({x \over {x+x_1}} )^2 \bigr \} ~ \biggr]
}
$

\noindent
$
{}~~~~~~~~~~~~~~~~~
\displaystyle{
- { K \over 6 } \int_x^1 {{dx_1} \over {x_1}}
         x_1 G(x_1) \sum_i^{\textstyle{-}}
         e_i^2 ~ \bigl \{x q_i^*(x) + x \bar q_i^*(x) \bigr \}
       ~  { x \over {x+x_1}} ~
         \bigl \{ 1+( {x \over {x+x_1}} )^2 \bigr \}
}
$
\hfill$(A2.3)$

$~~~~~$

\noindent
$
\displaystyle{
\Delta F_2^{(3)}(x) =
- { {4K} \over 9 } x \int_0^1 dx_2
         \sum_i^{\textstyle{-}} e_i^2 ~
    \bigl\{~ xq_i(x) x_2 \bar q_i^*(x_2)
                         +x\bar q_i(x) x_2 q_i^*(x_2)~ \bigr\} ~
         {{x^2+x_2^2} \over {(x+x_2)^4}}
}
$

\noindent
$
{}~~~~~~~~~~~~~~~~~
\displaystyle{
- { {4K} \over 9 } x \int_0^1 dx_2
         \sum_i^{\textstyle{-}} e_i^2 ~
       \bigl \{~   x q_i^* (x) x_2 \bar q_i(x_2)
                  +x \bar q_i^* (x) x_2 q_i(x_2)  ~\bigr\} ~
         {{x^2+x_2^2} \over {(x+x_2)^4}}
}
$
\hfill$(A2.4)$
\vskip 0.4cm

\noindent
where
$\displaystyle{\sum^{\textstyle{-}}}$ indicates
that the summation is averaged over partons in the proton
and the neutron. For example, they are
$\displaystyle{\sum_i^{\textstyle{-}}
 e_i^2 q_i = \bigl [ { 5 \over {18}} (u+d) + {2 \over {18}} s
             \bigr ]_{proton}
}$
and
$\displaystyle{\sum_i^{\textstyle{-}}
 e_i^2 q_i \bar q_i =
}$

\noindent
$\displaystyle{ \bigl [ { 5 \over {18}} (u \bar u+ d \bar d)
                    + {2 \over {18}} s \bar s
               \bigr ]_{proton}
}$.

In the similar way, modifications in
a gluon distribution in a nucleus are obtained as [\QIUCOM]

\vskip 0.4cm
\noindent
$
\displaystyle{
x\Delta G (x)
=x\Delta G^{(1)} (x)
+x\Delta G^{(2)} (x)
+x\Delta G^{(3)} (x)
}
$
\hfill$(A3.1)$

$~~~~~$

\noindent
$
\displaystyle{
x\Delta G^{(1)} (x) =
     +{ {3K} \over 4 } x \int_0^x dx_1 ~ x_1 G(x_1)~ (x-x_1) G^* (x-x_1)
}
$

$
\displaystyle{~~~~~~~~~~~~~~~~~~~~~~~~~~~
              ~~~~~~~~~~~~~
   ~ \times ~ { { 1 } \over {x^2}} ~
        \biggl\{  { {x_1} \over {x-x_1}} + {{x-x_1} \over {x_1}}
          +{{x_1(x-x_1)} \over {x^2}} \biggr\}
}
$

\noindent
$
{}~~~~~~~~~~~~~~~~~
\displaystyle{
     -{ {3K} \over 4 }  x \int_0^1 dx_2 ~
      ~x G (x)~ x_2 G^* (x_2)~
     { { 1 } \over { (x+x_2)^2 }}~
        \biggl\{  { x \over {x_2}} + {{x_2} \over x}
          +{{x x_2} \over { (x+x_2)^2} } \biggr\}
}
$

\noindent
$
{}~~~~~~~~~~~~~~~~~
\displaystyle{
     -{ {3K} \over 4 }  x \int_0^1 dx_2 ~
      ~x G^* (x)~ x_2 G (x_2)~
     { { 1 } \over { (x+x_2)^2 }}~
        \biggl\{  { x \over {x_2}} + {{x_2} \over x}
          +{{x x_2} \over { (x+x_2)^2} } \biggr\}
}
$
\hfill$(A3.2)$

$~~~$

\noindent
$
\displaystyle{
x \Delta G ^{(2)} (x) =
       +  { {4K} \over 9 } x \int_0^x dx_1
            \sum_i^{\textstyle{-}}
   ~  \bigl\{ ~ x_1 q_i(x_1) (x-x_1) \bar q_i^*(x-x_1)
}$

\noindent
$
\displaystyle{~~~~~~~~~~~~~~~~~~~~~
{}~+~ x_1 \bar q _i (x_1) (x-x_1) q_i^*(x-x_1) ~\bigr\} ~
           {1 \over {x^4}} ~
           \bigl \{  x_1^2 + (x-x_1)^2 \bigr \}
}
$
\hfill$(A3.3)$

$~~~~~$

\noindent
$
\displaystyle{
x \Delta G^{(3)} (x) =
- { K \over 6 }  \int_0^1 dx_2 ~x G(x)~
         \sum_i^{\textstyle{-}}
          \bigl\{ x_2 q_i^*(x_2) + x_2 \bar q_i^* (x_2) \bigr\}
       ~  {{1} \over {x+x_2}} ~
       \bigl\{ 1 + ( {{x_2} \over {x+x_2}} )^2 \bigr\}
}
$

\noindent
$
{}~~~~~~~~~~~~~~~~~
\displaystyle{
- { K \over 6 }  \int_0^1 dx_2 ~xG^*(x)~
         \sum_i^{\textstyle{-}}
        \bigl\{ x_2 q_i(x_2) + x_2 \bar q_i (x_2) \bigr\}
         {{1} \over {x+x_2}}~
       \bigl\{ 1 + ( {{x_2} \over {x+x_2}} )^2 \bigr\}
}
$
\hfill$(A3.4)$
\vskip 0.4cm

\noindent
where
$\displaystyle{\sum_i^{\textstyle{-}}
 q_i = \sum_i (q_i)_{proton}
}$
and
$\displaystyle{\sum_i^{\textstyle{-}}
 q_i \bar q_i = \sum_i (q_i \bar q_i)_{proton}
}$.
Using these expressions, we find that
the momentum conservation is explicitly satisfied,
$\displaystyle{
\int dx \bigl[ \sum_i x \bigl\{ \Delta q_i (x)
 + \Delta \bar q_i (x) \bigr\}
}$

\noindent
$\displaystyle{
          + x \Delta G (x) \bigr] ~=~ 0 .
}$
For numerical analysis, the last integral in $\Delta G^{(1)} (x)$
should be evaluated by separating the integral region
$\displaystyle{
\int_0^1 d x_2 = \int_0^x dx_2}$
$\displaystyle{+ \int_x^1 d x_2}$
and by changing variables $x'=x-x_2$ in the integral
$\displaystyle{\int_0^x dx_2}$. We use the following
equations for evaluating $\Delta G^{(1)}(x)$
in order to cancel out infinities:

\noindent
$\displaystyle{
x\Delta G^{(1)} (x)= \int_0^{x/2} dx' ~[~f(x,x')+f(x,x-x')~]
                    ~-~\int_x^1     dx' ~g(x,x')
}
$
\hfill$(A4.1)$
\vskip 0.4cm

$
\displaystyle{
 f(x,x') =
     +{ {3K} \over 4 } x ~ x' G(x')~ (x-x') G^* (x-x')
    ~ { { 1 } \over {x^2}} ~
        \biggl\{  { {x'} \over {x-x'}} + {{x-x'} \over {x'}}
          +{{x'(x-x')} \over {x^2}} \biggr\}
}
$

\noindent
$
{}~~~~~~~~~~~~~~~~~
\displaystyle{
     -{ {3K} \over 4 }  x  ~
      ~x G (x)~ x' G^* (x')~
     { { 1 } \over { (x+x')^2 }}~
        \biggl\{  { x \over {x'}} + {{x'} \over x}
          +{{x x'} \over { (x+x')^2} } \biggr\}
}
$

\noindent
$
{}~~~~~~~~~~~~~~~~~
\displaystyle{
     -{ {3K} \over 4 }  x ~
      ~x G^* (x)~ (x-x') G (x-x')~
     { { 1 } \over { (2x-x')^2 }}~
        \biggl\{  { {x} \over {x-x'}} + {{x-x'} \over x}
          +{{x (x-x')} \over { (2x-x')^2} } \biggr\}
}
$

\hfill$(A4.2)$

$
\displaystyle{
 g(x,x') =
     +{ {3K} \over 4 } x ~ [~ x G  (x)~ x' G^* (x') ~+~
                              x G^*(x)~ x' G   (x') ~]
}
$

\noindent
$
{}~~~~~~~~~~~~~~~~~~~~~~~~~~~~~~~~~~~~~~~~~~~~~
\displaystyle{
     \times
     { { 1 } \over { (x+x')^2 }}~
        \biggl\{  { x \over {x'}} + {{x'} \over x}
          +{{x x'} \over { (x+x')^2} } \biggr\}
}
$
\hfill$(A4.3)$

$~~~$

Nuclear gluon distributions in our model are discussed
in Ref. \SKGLUE ~
and they are compared with recent NMC measurements
of $G_{Sn}(x)/G_C (x)$.

\endpage
\par \penalty-400 \vskip\chapterskip
   \spacecheck\referenceminspace \immediate\closeout\referencewrite
   \referenceopenfalse
   \line{\fourteenrm\hfil References\hfil}\vskip\headskip
   \input referenc.tex
   
\endpage
\par \penalty-400 \vskip\chapterskip
   \spacecheck\referenceminspace
   \immediate\closeout\figurewrite \figureopenfalse
   \line{\fourteenrm\hfil Figure Captions\hfil}\vskip\headskip
   \input figures.aux
   
\bye
\end